\documentclass{aastex61}
\usepackage{amsmath}

\shorttitle{Distance and Metallicity Gradients Along Andromeda's GSS From the Red Clump}
\shortauthors{R.E.~Cohen et al.}
\begin{document}
\title{Project AMIGA: Distance and Metallicity Gradients Along Andromeda's Giant Southern Stream from the Red Clump\footnote{Based on observations made with the NASA/ESA Hubble Space Telescope, obtained at the Space Telescope Science Institute, which is operated by the Association of Universities for Research in Astronomy, Inc., under NASA contract NAS 5-26555.  These observations are associated with program GO-14268.}}

\correspondingauthor{Roger E. Cohen}
\email{rcohen@stsci.edu}

\author{Roger E. Cohen}
\affiliation{Space Telescope Science Institute, 3700 San Martin Drive, Baltimore, MD 21218, USA}

\author{Jason S. Kalirai}
\affiliation{Space Telescope Science Institute, 3700 San Martin Drive, Baltimore, MD 21218, USA}

\author{Karoline M. Gilbert}
\affiliation{Space Telescope Science Institute, 3700 San Martin Drive, Baltimore, MD 21218, USA}
\affiliation{Department of Physics and Astronomy, The Johns Hopkins University, 3400 N.~Charles Street, Baltimore, MD 21218, USA}

\author{Puragra Guhathakurta}
\affiliation{UCO/Lick Observatory, Department of Astronomy \& Astrophysics, University of California Santa Cruz, 1156 High Street, Santa Cruz, CA 95064, USA}

\author{Molly S. Peeples}
\affiliation{Space Telescope Science Institute, 3700 San Martin Drive, Baltimore, MD 21218, USA}

\author{Nicolas Lehner}
\affiliation{Department of Physics, University of Notre Dame, 225 Nieuwland Science Hall, Notre Dame, IN 46556, USA}

\author{Thomas M. Brown}
\affiliation{Space Telescope Science Institute, 3700 San Martin Drive, Baltimore, MD 21218, USA}

\author{Luciana Bianchi}
\affiliation{Department of Physics and Astronomy, The Johns Hopkins University, 3400 N.~Charles Street, Baltimore, MD 21218, USA}

\author{Kathleen A. Barger}
\affiliation{Department of Physics and Astronomy, Texas Christian University, TCU Box 298840, Fort Worth, TX 76129, USA}

\author{John M. O'Meara}
\affiliation{Department of Chemistry and Physics, Saint Michael's College, One Winooski Park, Colchester, VT 05439, USA}

\begin{abstract}

The Giant Southern Stream (GSS) of M31, a keystone signature of a major accretion event, yields crucial constraints on M31 formation and evolution models.  Currently, our understanding of the GSS, in terms of both its geometry and its chemistry, results from either wide-field imaging probing only a few magnitudes below the red giant branch tip, or deep 
imaging or spectroscopy of isolated regions.  Here, we take an alternative approach, using \textit{Hubble Space Telescope} (HST) imaging to characterize the horizontal branch red clump (RC) using unbinned maximum likelihood fits to luminosity functions (LFs) from observed color-magnitude diagrams (CMDs).  
Comparing the RC mean magnitude across three fields at projected distances of 21, 52 and 80 kpc from M31, we find a line of sight distance gradient identical to recent literature measurements in fields along the core.
We also find tentative evidence that the line of sight distance dispersion increases with projected distance from M31.  
Meanwhile, the metallicity in the 52 kpc field westward of the GSS core is at least as high as that in the 21 kpc GSS core field, and the peak colors of the RC in these two fields imply identical metallicities to within 0.2 dex.  We discuss implications for distance and metallicity gradients both along and perpendicular to the GSS in the context of recent ground-based photometric and spectroscopic results, including evidence for a dropoff in metallicity moving westward from the GSS, as well as prospects for further constraining stellar populations in the vicinity of the GSS.

\end{abstract}

\section{Introduction} \label{sec:intro}

Andromeda's Giant Southern Stream (or Giant Southern Stream, hereafter GSS) is an archetypal signature of a major accretion event seen in halo substructure.  Since its discovery from wide-field ground-based imaging \citep{ibata01,ferguson02}, 
subsequent ground-based surveys have revealed that it is a huge coherent structure: The GSS extends across 6 degrees of the sky to at least $\sim$80 kpc in projected radius $R_{\text{proj}}$, 
over which it spans a roughly linear line-of-sight distance gradient of $\sim$100 kpc \citep{m03,conn} accompanied by a radial velocity gradient of $>$200 km/s \citep{ibataspec,gilbert_gss,gilbert_new}.  While the GSS is generally both more metal-rich \citep{ferguson02,raja06,gilbert_feh_global} and kinematically colder \citep[e.g.][]{raja06} than the surrounding halo at fixed $R_{\text{proj}}$, a more complex picture is being revealed both chemically and kinematically.  With regard to metallicity, imaging \citep{ibata07,ibataglobal} and spectrocopy \citep{gilbert_gss} both illustrate that the GSS core is more metal-rich than the envelope (also see \citealt{koch_gas}).  Regarding kinematics, the velocity distribution of the GSS population has multiple cold peaks in both the GSS core \citep{jason_gss_spec} and the envelope \citep{gilbert_gss}.  

The above ensemble of positional and kinematic information has been leveraged together in models in a hunt for the nature and present day location of the GSS progenitor.  Since initial predictions by \citet{ibataspec} and \citet{font_model}, remarkable progress has been made in modeling the velocity and density distributions of the GSS by \citet{fardal_nbody,fardal_disk}, arguing in favor of a cold, rotating disky progenitor.  Subsequently, \citet{fardal_m31mass} used a Plummer sphere progenitor while estimating the mass of the GSS as similar to that of the Large Magellanic Cloud and predicting the mass of M31.  Focusing on the nature of a disky progenitor, \citet{kirihara} simulate disrupting infalling disks (with and without a bulge) over a range of initial parameters including axis orientation relative to the M31 disk and progenitor metallicity gradient, and find sets of initial conditions that reproduce the density and/or metallicity gradients in the GSS, also yielding predictions for the azimuthal metallicity gradient across the GSS at various $R_{\text{proj}}$.  Similarly, predictions for the line-of-sight distribution of GSS material have recently been made by \citet{m31_recentsims} arguing for the paradigm of a major merger event in the last few Gyr.  

Observational constraints on such model predictions have thus far been mostly limited to two categories.  Wide-field surveys have the advantage of homogeneity, but probe only a few magnitudes below the tip of the red giant branch (TRGB) and therefore rely on the foreground-decontaminated TRGB for distance measurements \citep[e.g.][]{m03,conn}.  Conversely, deep ground-based spectroscopic \citep{ibataspec,gilbert_gss,gilbert_feh_global,gilbert_new} and space-based photometric campaigns \citep{ferguson05,brown_streamvsph, browncats} yield detailed information on abundances, kinematics and/or star formation histories, but only for a set of isolated, \textquotedblleft pencil-beam\textquotedblright{} locations.  These studies have provided a plethora of information on their target fields, especially where spectroscopic and deep photometric fields overlap.  We now know that the GSS at $R_{\text{proj}}$=21 kpc has a star formation history that is strikingly similar, but not identical, to a minor axis halo field at $R_{\text{proj}}$=11 kpc, such that both host populations with extended distributions of both metallicity ($>$2 dex) and age ($>$10 Gyr), although the GSS field is on average about 1 Gyr younger, with a mean age of $\sim$8-9 Gyr and a mean metallicity of $[\rm{Fe/H}]$$\sim$-0.5 dex \citep{brownsfh, bernard_sfh}.

Stellar population gradients perpendicular to the GSS remained observationally unconstrained until several years after the discovery of the GSS.  \citet{ibata07} used contiguous imaging of the upper RGB over a large survey area to demonstrate the difference in metallicity between the GSS core and its envelope to the west, such that the core contains a higher fraction of metal-rich stars.  The core-envelope difference was underscored spectrocopically by \citet{gilbert_gss}, confirming a decrease of at least $\sim$0.4 dex in metallicity in the envelope and revealing the presence in the envelope of kinematic substructure previously found only in the GSS core.  Global metallicity maps of Andromeda's halo \citep{ibataglobal} appear inconclusive as to the sharpness of a metallicity dropoff moving westward from the GSS core to the envelope, while illustrating that the mean metallicity in the GSS core remains high over much of its extent, revealing [\rm{Fe/H}]$\gtrsim$-0.5 out to $R_{\text{proj}}$$\sim$60 kpc.

Now that models are providing specific, testable predictions regarding distance, age and metallicity gradients not just along the GSS, but also perpendicular to it azimuthally and along the line of sight, we employ a different strategy to constrain stellar population gradients from deep imaging.  We measure the properties of the horizontal branch red clump (RC) in three fields along the GSS with 21$\leq$$R_{\text{proj}}$$\leq$80 kpc, imaged with an identical instrument and filter setup.  In this way, we exploit an available lever arm of $\sim$60 kpc in $R_{\text{proj}}$ to measure gradients in distance, distance distribution, and metallicity.  Although the use of the RC as a standard candle or standard crayon comes with several important caveats (discussed in Sect.~\ref{standardcandlesect}, see \citealt{girardirc} for a review), we use a strictly differential analysis to minimize the impact of absolute assumptions on stellar astrophysics inherent in evolutionary models.    

In Sect.~\ref{obssect} we describe the observations and photometry.  In Sect.~\ref{anasect}, we measure
the properties of the red clump and horizontal branch, and the metallicity distribution of the upper RGB
in each of our target fields.  In Sect.~\ref{discusssect} we discuss the constraints on GSS distance,
metallicity and age gradients resulting from our analysis and compare them with recent literature results,
and in the last section we summarize our results.

\section{Data} \label{obssect}
\subsection{Observations}

\begin{figure}
\gridline{\fig{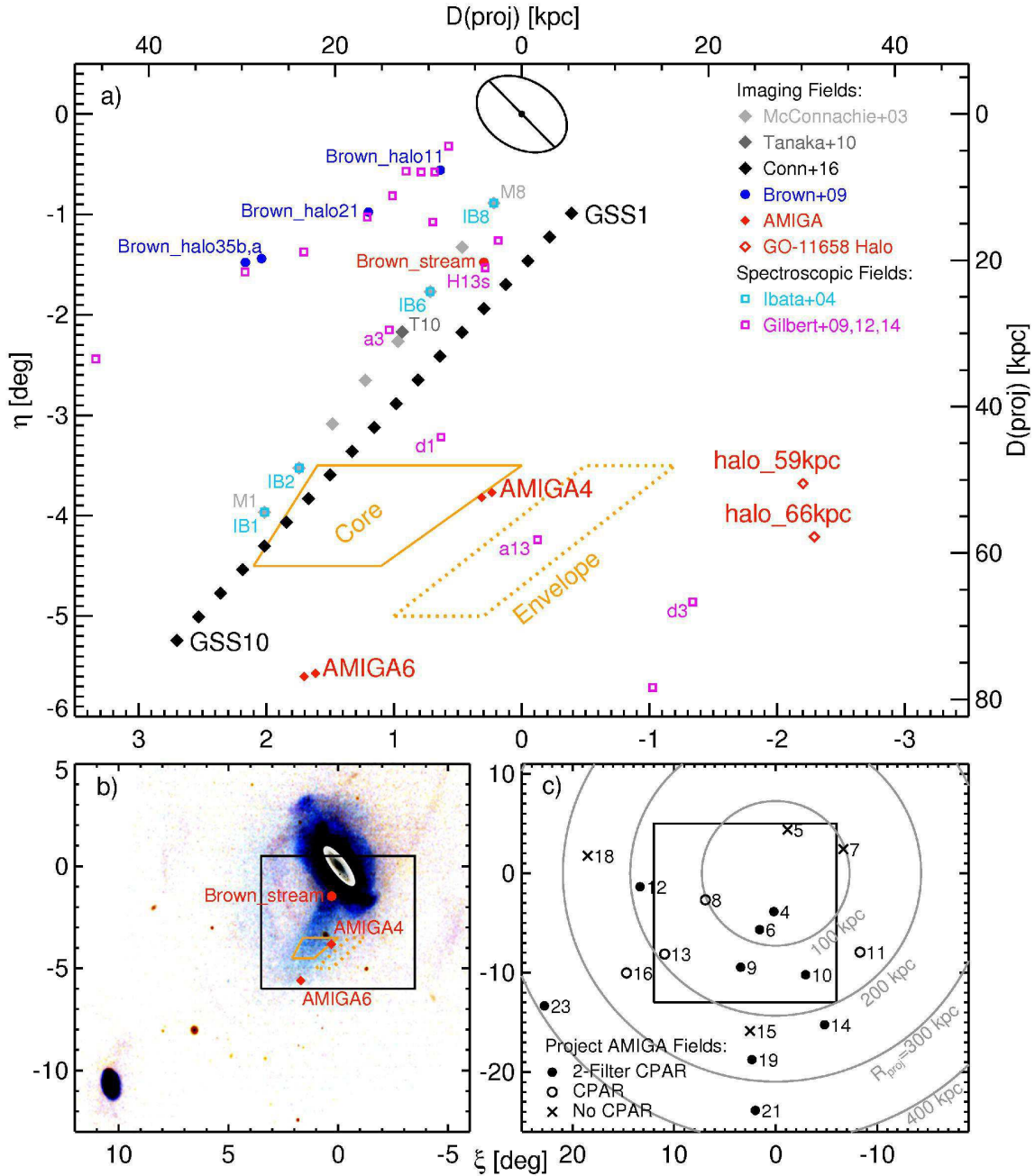}{0.85\textwidth}{}
	 } 
\caption{\textbf{(a,top):} Map of existing imaging and spectroscopy in the vicinity of the GSS, shown in M31-centric coordinates (north is up and east is to the left).  The GSS core and envelope regions from \citet{ibata07} are indicated by solid and dotted orange polygons respectively, and the disk of M31 is depicted using structural parameters from \citet{dorman_m31disk}.  Previously analyzed ground-based imaging is shown in greyscale, including fields M1-M8 from \citet{m03}, fields GSS1-GSS10 from \citet{conn} and the SuprimeCam field analyzed by \citet{tanaka}.  The HST ACS/WFC fields from \citet{browncats} are shown as circles, the Project AMIGA GSS fields we analyze here are shown as filled red diamonds (corresponding to the individual locations of the ACS/WFC and WFC3/UVIS pointings), and M31 halo fields with archival ACS/WFC imaging used to measure M31 halo contamination in our AMIGA GSS fields (see Sect.~\ref{halocontamsect}) are shown as open red diamonds.  Spectroscopic fields are shown as open squares, with fields from the SPLASH survey \citep{gilbert_gss,gilbert_sb_global,gilbert_feh_global} in magenta and fields with radial velocities from \citet{ibataspec} in cyan. \textbf{(b,lower left):} A PAndAS density map from \citet{martin_pandas_densmap} with color rescaled to highlight the GSS.  Approximate locations of our three target fields are overplotted in red, as well as the locations of the core and envelope regions as in panel (a), and the location of panel (a) is indicated by a black box.  \textbf{(c, lower right):} Map of all of the Project AMIGA fields, with field number indicated.  Different symbols correspond to fields with contiguous parallel (CPAR) imaging in two filters (solid circles), fields where parallel imaging in the two filters is not spatially coincident (open circles), and fields with no parallel imaging (crosses). The location of panel (b) is shown as a box. \label{posfig}}
\end{figure}

The observations we analyze consist of \textit{Hubble Space Telescope (HST)} 
coordinated parallel imaging obtained as part of Project AMIGA
(GO-14268, PI:Lehner).  The primary science driver of Project AMIGA 
(Absorption Maps in the Gas of Andromeda) is to leverage the diagnostic power of ultraviolet COS\footnote{Cosmic Origins Spectrograph} 
spectra along 
an ensemble of QSO sightlines through Andromeda's halo, probing the 
circumgalactic medium at projected radii
of 25$<$$R_{\text{proj}}$$<$339 kpc from M31.  During the COS primary exposures along
each QSO sightline,
parallel imaging was obtained simultaneously 
with both ACS/WFC and WFC3/UVIS in the
$F606W$ and $F814W$ filter of each instrument when feasible, yielding two-color imaging of 
two pointings separated by $\sim$5.8 arcmin for each AMIGA field.
Because the placement and readout of the primary COS exposures
constrained these parallel images 
in depth, observing cadence, roll angle and dither pattern, parallel imaging was not obtained for some QSO sightlines in one or both filters or instruments.   Here we focus on the two innermost AMIGA fields which do have two-filter parallel imaging,
denoted AMIGA4 and AMIGA6, because their CMDs and LFs show clear evidence of a horizontal branch red clump (RC). These two fields are fortuitously located close to the GSS at projected radii of $R_{\text{proj}}$=52 and 80 kpc from M31 (see Fig.~\ref{posfig}) assuming $(m-M)_{0,M31}$ = 24.47 \citep{m31dist}.  Therefore, by combining these fields with archival ACS/WFC $F606W,F814W$ imaging and photometry of a GSS field at $R_{\text{proj}}$=21 kpc, which we denote Brown\_stream, 
we can directly compare the RC color and magnitude distributions of the three fields to place independent constraints
on distance and metallicity gradients along the GSS over a baseline of $\Delta$$R_{\text{proj}}$$\sim$60 kpc.  By effectively using the 21 kpc Brown\_stream field as a baseline for comparison to the AMIGA4 and AMIGA6 fields, we take advantage of the known star formation history (i.e.~age and metallicity distributions) of the 21 kpc field, along with its similarity to an 11 kpc minor axis halo field \citep{brown_streamvsph,brownsfh,browncats,bernard_sfh}. 
In Table \ref{targtab}, we give the locations,
number of exposures and total exposure time per filter per instrument for each of the target fields analyzed here.  

\begin{deluxetable}{lcccccccccc}
\tablecaption{Target GSS Fields \label{targtab}}
\tablehead{
\colhead{Field} & \colhead{$R_{\text{proj}}$} & \colhead{RA (J2000.0)} & \colhead{Dec (J2000.0)} & \colhead{$\xi$} & \colhead{$\eta$} & \colhead{Instrument} & \colhead{$N_{exp}$} & \colhead{$t_{exp}$} & \colhead{$N_{exp}$} & \colhead{$t_{exp}$} \\ \colhead{} & \colhead{kpc} & \colhead{$^\circ$} & \colhead{$^\circ$} & \colhead{$^\circ$} & \colhead{$^\circ$} & \colhead{} & \multicolumn{2}{c}{$F606W$} & \multicolumn{2}{c}{$F814W$}
}
\startdata
AMIGA4 & 52 & 11.09075 & 37.45003 & 0.3139 & -3.8187 & ACS/WFC & 7 & 4193 & 9 & 5580 \\
       & 52 & 10.98718 & 37.49975 & 0.2338 & -3.7690 & WFC3/UVIS & 6 & 4035 & 6 & 4535 \\
AMIGA6 & 79 & 12.82040 & 35.71136 & 1.6716 & -5.5574 & ACS/WFC & 8 & 3174 & 8 & 3920 \\
       & 80 & 12.70254 & 35.70301 & 1.5795 & -5.5657 & WFC3/UVIS & 4 & 1810 & 4 & 2220 \\
Brown\_stream & 21 & 11.07500 & 39.79222 & 0.2966 & -1.4765 & ACS/WFC & 4 & 5060 & 4 & 5060 \\  
\enddata
\end{deluxetable}

\subsection{Preprocessing and Photometry \label{photsect}}

We retrieved individual preprocessed \texttt{.flc} science images as well as a stacked, drizzled, distortion-corrected \texttt{.drc} reference image from the \textit{HST} archive for each pointing.  
For subsequent preprocessing and photometry of all images for each pointing, we used version 2.0 of the publicly available \texttt{Dolphot} package\footnote{\url{http://americano.dolphinsim.com/dolphot/}} \citep{dolphin}.  
We performed additional preprocessing according to the recommendations of each
(instrument-specific) \texttt{Dolphot} manual, including masking bad pixels, separating each science image into individual chips for photometry, and calculating the sky background\footnote{We use the high-resolution \texttt{step}=4 value for generating sky frames.}.

\texttt{Dolphot} uses customized PSFs tailored to each filter of each \textit{HST} instrument to perform iterative PSF photometry simultaneously across multiple science images which are aligned to the deep reference image.  
Many optional parameters governing how image alignment, PSF fitting and sky subtraction are performed may be altered within \texttt{Dolphot}, and in cases of severe stellar crowding, modifying these parameters can result in cleaner, more complete photometric catalogs \citep{williamsphat}.  However, given the low stellar densities in our target fields, we found that the ability to identify and reject background galaxies was a much stronger determinant of photometric quality than stellar crowding.  In particular, rather than setting \texttt{Force1}=1, which forces all objects to be classified as stellar and can improve photometric precision in crowded fields,
we retain \texttt{Force1}=0 to exploit the ability of \texttt{Dolphot} to flag suspected background galaxies, although some non-zero galaxy contamination still remains (see Sect.~\ref{contamsect}).  Since the ACS/WFC images retrieved from the archive were already corrected for charge transfer efficiency (CTE) by the \texttt{calacs} pipeline, we set \texttt{UseCTE}=0.
For each set of input images, \texttt{Dolphot} outputs a photometric catalog calibrated to the Vegamag system using the encircled energy corrections and photometric zeropoints from \citet{bohlin} for ACS/WFC and from \citet{wfc3} 
for WFC3/UVIS.

The WFC3/UVIS catalogs were concatenated with the ACS/WFC catalogs in order to optimize number statistics in the AMIGA4 and AMIGA6 fields,
which are relatively sparsely populated compared to fields at smaller projected distances from M31.  Publicly available programs from \citet{cv14} were used to calculate linear transformation coefficients converting WFC3/UVIS $F606W,F814W$ magnitudes to the $F606W,F814W$ in the ACS/WFC photometric system\footnote{These transformations take into account the shift from current WFC3/UVIS zeropoints, valid for our observations, and pre-2016 WFC3/UVIS zeropoints used by \citet{cv14}}.  The coefficients were derived using synthetic
stellar photometry of 12 Gyr stars with -2$\leq$[\rm{Fe/H}]$\leq$0 over the full range of log g covered by their models, noting that variations due to changes in stellar parameters (i.e.~source spectral energy distribution) are predicted to be $<$0.01 mag for $(F606W-F814W)_{0}$$<$3 but increase to 0.02 mag for metal-rich red giants near the TRGB.  
For the 21 kpc GSS core field, photometric catalogs and artificial star tests were made publicly available in \citet{browncats}, but we analyze photometry from a \texttt{Dolphot} re-reduction as described above for three reasons.  First, we avoid systematic errors due to zeropoint uncertainties in converting between the STmag and Vegamag photometric systems.  Second, we can directly apply our background galaxy contamination maps from imaging processed identically using \texttt{Dolphot} (see Sect.~\ref{galcontamsect}).  Third, we have intentionally reprocessed only a subset of the imaging analyzed by \citet{browncats} in order attain a similar photometric depth and precision as in the two AMIGA fields.  This eliminates substantial differences in photometric quality as a potential cause of differences in the RC magnitude dispersion across our fields (see Sect.~\ref{rcmagdispersionsect}).

Artificial star tests were performed, also using \texttt{Dolphot}, to quantify incompleteness and photometric errors in our catalogs.  For each field, two sets of artificial star tests are performed.  The stars are assigned a flat LF in $F814W$ in both sets of tests, but in one set stars are assigned colors randomly drawn from the observed color distribution at their magnitude, and in the other set, stars are gridded evenly over the entire color-magnitude space occupied by the observed catalogs.  By inserting $>$10$^{5}$ artificial stars per field (over 10 times the number in any of the observed catalogs) one at a time, we build up a sufficient sample to quantify the distribution in color and magnitude error at any CMD location (analogous to the scattering kernels of \citealt{browncats}), as well as the variation of completeness over the CMD.  A large ensemble of artificial stars also reveals the regions in diagnostic parameter space (i.e.~\texttt{chi} or \texttt{sharp} versus magnitude) where only spurious detections are expected to lie, since they will be occupied exclusively by sources in the raw observed catalog but not the catalog of recovered values for input artificial stars.  With this in mind, we found that an optimal compromise between eliminating spurious detections and background galaxies versus photometric completeness was obtained by imposing a set of cuts similar to \citet{martindolphotcuts}.  Specifically, we only retained sources from our 
observed catalogs with $\sqrt{\texttt{sharp}_{F606W}^{2}+\texttt{sharp}_{F814W}^{2}}$$\leq$0.1, $(\texttt{crowd}_{F606W}+\texttt{crowd}_{F814W})$$\leq$1, reported S/N$\geq$5 per filter in both filters, a photometric quality flag $\leq$2 and an object type of 1 (suspected non-stellar sources are flagged by object types greater than 1).  Over the color range occupied by GSS RGB and RC stars, 
the resulting catalogs are $>$90\% complete from brightward of the M31 TRGB to faintward of the RGB bump and completeness varies slowly and smoothly as a function of color and magnitude\footnote{Because the sensitivity of WFC3/UVIS is different than that of ACS/WFC, artificial star tests were performed separately for the two instruments, and the area-weighted results were concatenated to measure completeness and photometric errors as a function of CMD location for each field.}.

\begin{figure}
\gridline{\fig{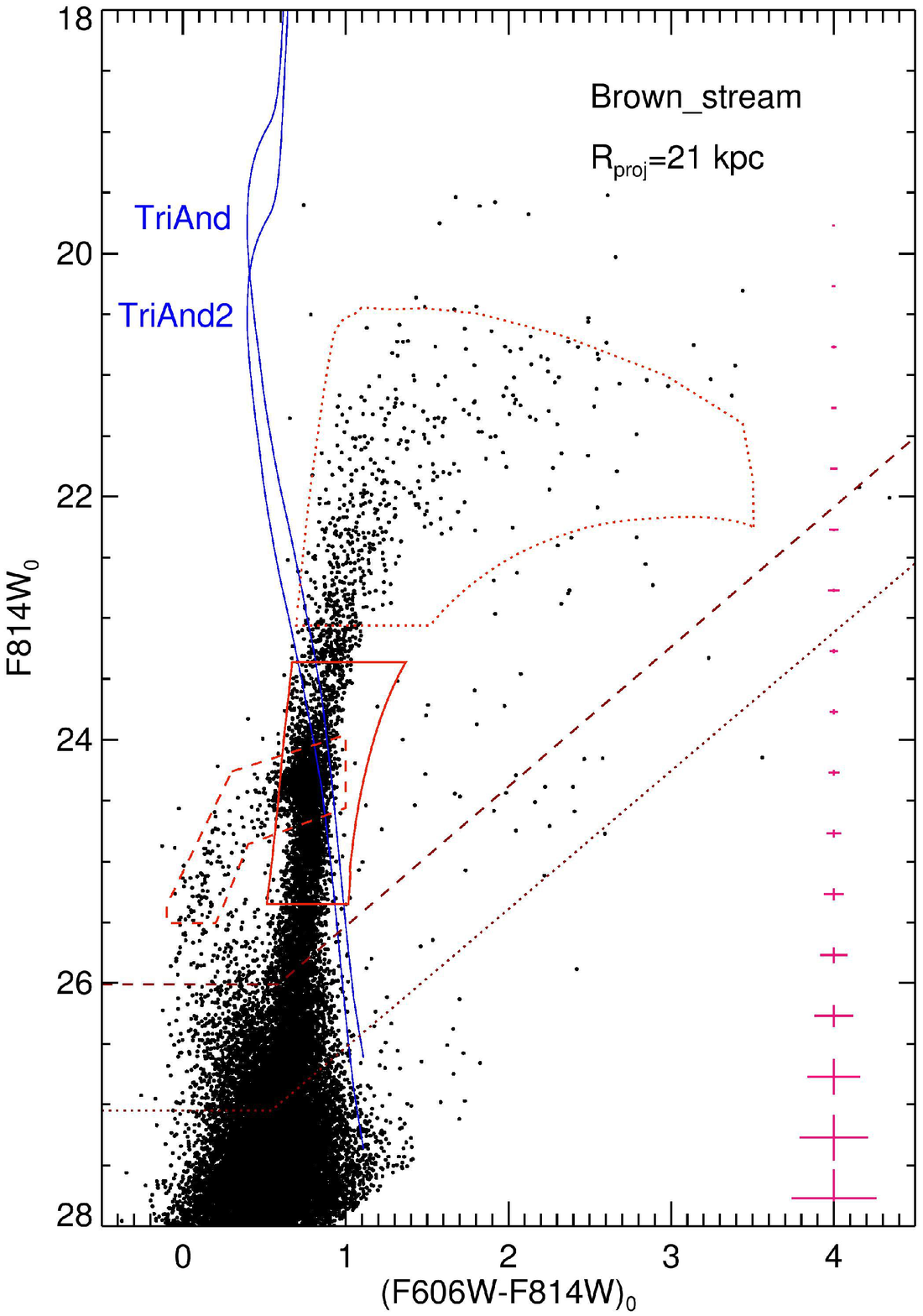}{0.33\textwidth}{}
	  \fig{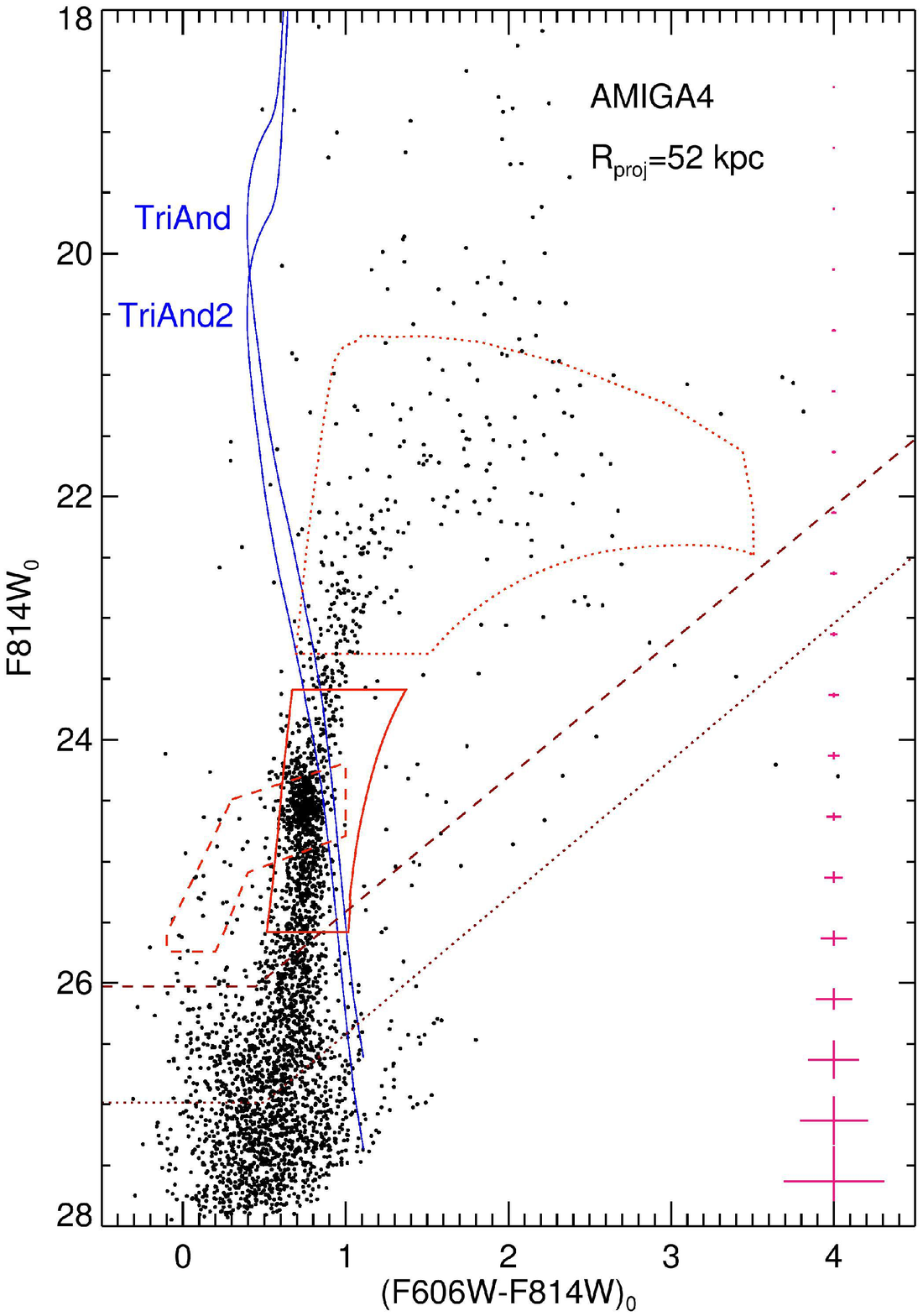}{0.33\textwidth}{}
	  \fig{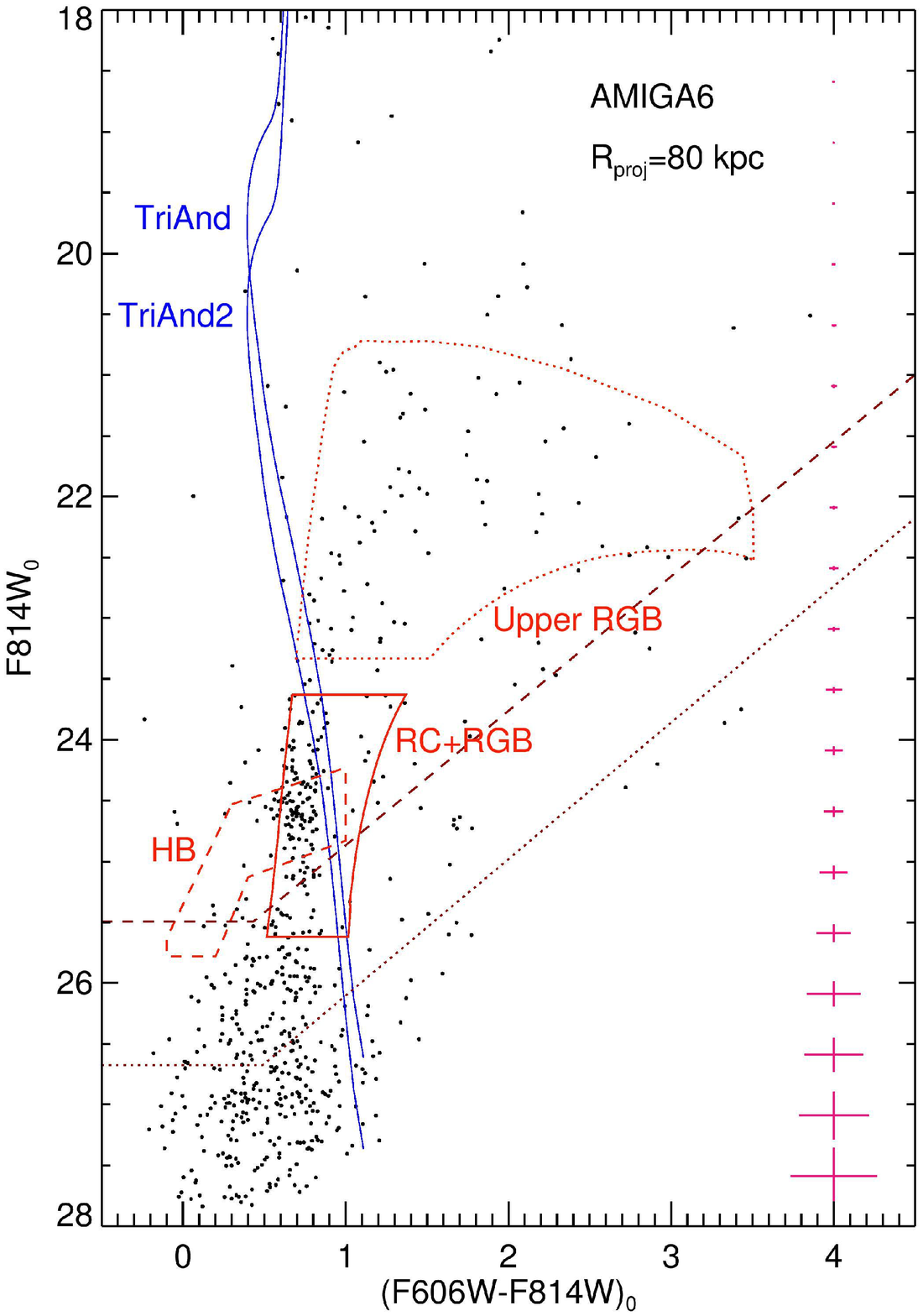}{0.33\textwidth}{}}
\caption{CMDs of our target fields, illustrating regions used to select RC+RGB stars (solid red line), horizontal branch stars (dotted red line) and upper RGB stars used to obtain photometric metallicity distributions (dotted red line).  Isochrones representing the CMD location of the TriAnd and TriAnd2 Milky Way halo overdensities \citep{martin_triand2} are shown in blue.  The 90\% and 50\% completeness limits ascertained from artificial star tests are shown as dashed and dotted brown lines respectively, and photometric errors are indicated along the right hand side of each CMD. \label{cmdfig}}
\end{figure}

For each field, the photometric catalogs were corrected for foreground extinction using the \citet{schlafly} recalibration of the
\citet{sfd} reddening maps and the extinction coefficients given in \citet[][their table A1]{cv14}.  These maps reveal low foreground extinction of $E(B-V)<0.05$ for all of the sightlines analyzed here.  CMDs of each of our target fields are shown in Fig.~\ref{cmdfig}, and all of the magnitudes and colors we report are in the ACS/WFC Vegamag system, corrected for foreground extinction.

\section{Photometric Constraints on Stellar Populations Along the Stream \label{anasect}}

Our goal is to use observed color-magnitude diagrams (CMDs) to constrain age, distance and metallicity gradients along the GSS.  
While the sizes of the ACS/WFC and WFC3/UVIS fields are limiting in terms of number statistics, we may exploit their
depth and homogeneity to perform a strictly differential comparisons between observed properties of the GSS at $R_{\text{proj}}$=21, 52 and 80 kpc.  These comparisons are made using the following metrics:

\begin{enumerate}

\item{Observed magnitude and width of the red clump in both filters.}

\item{Observed color of the red clump.}

\item{Observed color distribution of the horizontal branch (HB).}

\item{Photometric metallicity distribution functions (MDFs) of upper RGB stars.}

\end{enumerate}

\subsection{Red Clump Magnitude \label{rcmagsect}}

We measure the magnitude and width of the RC in both filters using unbinned maximum likelihood fits to the LF in each filter.  The LF is constructed from a restricted region in the CMD, shown using a solid red line Fig.~\ref{cmdfig}, to minimize contamination of the LF by foreground Milky Way stars and unresolved background galaxies (see Sect.~\ref{contamsect}).  This CMD region, which we refer to as the RC+RGB region, is generated using a two-step process as follows:  First, a rough guess for the RC magnitude is calculated using binned histograms of stars with 0.5$\leq$$(F606W-F814W)_{0}$$\leq$1.0 and 23$\leq$$F814W_{0}$$\leq$26, shifted by a random fraction of the binsize over 1000 monte carlo iterations.  The location of the LF maximum averaged over all of the iterations is taken as an initial guess for the RC magnitude.  Next, the RC+RGB region is defined using 10 Gyr solar-scaled Dartmouth isochrones \citep{dsed}, selected to allow a direct comparison with recent literature studies (see Sect.~\ref{litcompsect}),
and we note that current spectroscopic measurements of [$\alpha$/Fe] in the halo of M31 remain inconclusive \citep{afe}.

The region defined by the resulting set of isochrones was then broadened in color using the 1-$\sigma$ photometric errors resulting from the artificial star tests in both AMIGA fields.  This region is then shifted faintward making the approximation that the difference in the rough guess $F814W_{0}$ magnitude of the RC between each field and the 11 kpc minor axis field from \citet{brown_streamvsph,browncats} is due to a difference in line-of-sight distance, and assuming the 11 kpc minor axis field lies at the same distance as M31.  
In other words, an identical RC+RGB CMD selection region is used for each field, but with vertical shifts applied corresponding to the shifts in rough guess RC magnitude in $F814W_{0}$.  To minimize contamination from both foreground halo sources and background galaxies, we further restrict the RC+RGB region to stars lying within $\pm$1 mag of the rough guess HB magnitude.  This region is illustrated as a solid red line in Fig.~\ref{cmdfig}.

\begin{figure}
\plotone{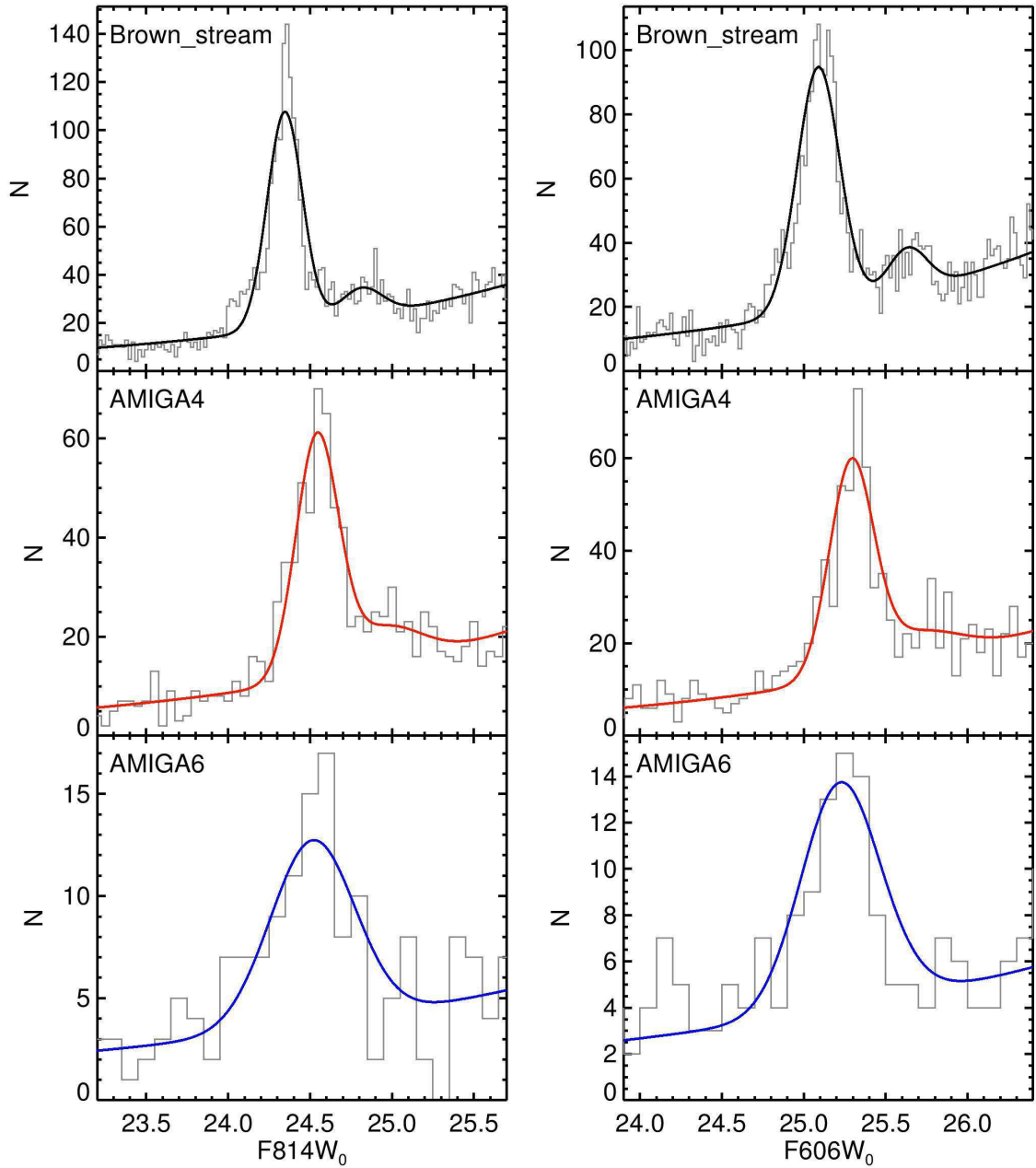}
\caption{Luminosity functions for each of our target fields in $F814W_{0}$ (left column) and $F606W_{0}$ (right column).  To facilitate a visual comparison, observed luminosity functions are shown using bin sizes of 0.02, 0.05 and 0.1 mag, but our maximum likelihood fits were performed \textit{without} binning the data.  The best fitting exponential plus double Gaussian functions as determined by the maximum likelihood fits are shown in each panel, color-coded by target field. \label{lfobsfig}}
\end{figure}

The observed LFs are shown for each target field in Fig.~\ref{lfobsfig}.  For the Brown\_stream and AMIGA4 fields, an excess likely representative of the red giant branch bump (RGBB) is seen slightly faintward of the RC (also noted by \citealt{brown_streamvsph} in the 21 kpc field).  
Because the RGBB has a strength (relative to the underlying RGB) and a peak magnitude (relative to the HB) which are correlated with metallicity, analytical template functions which employ two Gaussians to characterize both the RC and the RGBB  \citep[e.g.][]{natafrgbb} may provide valuable clues regarding the similarity between the Brown\_stream and AMIGA4 fields.  Therefore, all sources passing our photometric quality cuts and lying inside the RC+RGB CMD selection region are fit with an exponential plus double Gaussian analytical template function to represent the differential LF:

\begin{equation}
N(m) = A \Bigg\{\exp \big[ B(m-m_{RC})\big] + \frac{EW_{RC}}{\sqrt{2\pi}\sigma_{RC}} \exp \frac{-(m-m_{RC})^{2}}{2\sigma_{RC}^{2}} + \frac{EW_{RGBB}}{\sqrt{2\pi}\sigma_{RGBB}} \exp \frac{-(m-m_{RGBB})^{2}}{2\sigma_{RGBB}^{2}}   \Bigg \} 
\label{eq_dblgauss}
\end{equation}

The coefficient $A$ is a normalization constant and the first term represents the underlying RGB with exponential slope $B$.  The second term represents the RC as a Gaussian with an area $EW_{RC}$ relative to the underlying RGB, a standard deviation $\sigma_{RC}$ 
and mean magnitude $m_{RC}$.  Similarly, the third term is a Gaussian representing the RGBB, with a mean magnitude $m_{RGBB}$, a standard deviation $\sigma_{RGBB}$ and an area relative to the underlying RGB of $EW_{RGBB}$.  The number of RC and RGBB stars, $N(RC)$ and $N(RGBB)$, can then be calculated directly since $EW_{RC} = N(RC)/A$ and $EW_{RGBB} = N(RGBB)/A$.

The stellar densities of our target fields vary by a factor of $\sim$50 (see below), so an unbinned maximum likelihood approach is crucial to enable a consistent comparison of the RC (and RGBB) parameters and their uncertainties among our target fields. 
Following \citet{natafrgbb}, the log-likelihood $\ell$ of the distribution in Eq.~\ref{eq_dblgauss} over $N_{obs}$ data points in filter $m$ can be calculated as:

\begin{equation}
\ell_{m} = \sum_{i}^{N_{obs}} \ln \big[ N(m_{i}/A, B, EW_{RC}, \sigma_{RC}, m_{RC}, EW_{RGBB}, \sigma_{RGBB}, m_{RGBB})  \big] - N_{obs}
\label{eq_gaussian_maxl}
\end{equation}

We use the \texttt{emcee} \citep{emcee} implementation of the \citet{gwmcmc} Markov Chain Monte Carlo (MCMC) sampler to fit for the the LFs simultaneously in both filters by maximizing the total log-likelihood, which is the sum of the log-likelihoods in the two individual filters, each given by Eq.~\ref{eq_gaussian_maxl}:

\begin{equation}
\ell = \ell_{F606W} + \ell_{F814W}
\end{equation}

The success of this procedure requires that the normalization constant $A$ is expressed analytically as a function of the other parameters in each MCMC iteration, accomplished by setting the total area under Eq.~\ref{eq_dblgauss} over the sampled magnitude range equal to $N_{obs}$.  This area is just the sum of the area under the two Gaussians, $EW_{RC}$ and $EW_{RGBB}$, plus the finite integral of the exponential function representing the RGB over the sampled magnitude range.

In general, only generous flat priors were imposed on the distributions for the RC parameters: We required 0$<$$B$$\leq$1, 0$<$$EW_{RC}$$\leq$3, 0$<$$\sigma_{RC}$$\leq$1 and $m_{RC}$ to be within $\pm$1 mag of the rough guess value\footnote{Flat priors are not strictly uninformative (for example, they are not scale-invariant), but tests using priors which were more or less restrictive by a factor of two yielded results in agreement with those reported here to at least the 90\% confidence level.}.  While fitting in both filters simultaneously has the advantage that statistically valid results and uncertainties are obtained for quantities involving both filters (for example, the RC color $V(RC)-I(RC)$ discussed in Sect.~\ref{rccolorsect}), it comes at the cost of a large number of free parameters.  Therefore, to avoid potential degeneracies between RC and RGBB properties, we impose several additional astrophysically motivated priors, based on values observed for Galactic globular clusters over a wide range of metallicities (-2$\lesssim$$[\rm{Fe/H}$$\lesssim$0) from \citet{natafrgbb}:

\begin{enumerate}

\item The RGB slope $B$ is required to be the same in both filters.

\item $\sigma_{RGBB}$$<$0.5 in both filters.

\item The number ratio of RGBB to RC stars $f^{RC}_{RGBB} = N(RGBB)/N(RC)$ is required to be the same in both filters, and is restricted to values of $f^{RC}_{RGBB}$$<$0.6.

\item We require a magnitude difference between the RC and RGBB of -0.2$<$$m^{RC}_{RGBB}$$<$+0.8, which is conservative given the photometric metallicity distributions towards our target fields (see Sect.~\ref{photmetsect}).  In addition, the values of $m^{RC}_{RGBB}$ in each filter are required to be within $\pm$0.2 mag of each other, so that $\lvert$$V^{RC}_{RGBB}-I^{RC}_{RGBB}$$\rvert$$<$0.2.  Importantly, we do not \textit{require} that $V^{RC}_{RGBB}=I^{RC}_{RGBB}$ since doing so could artificially impose values on the RC color that we wish to measure.  However, placing a conservative upper limit on the difference in values merely serves to eliminate unphysical cases of a significant RGBB detection (in terms of $EW_{RC}$ and $m^{RC}_{RGBB}$) in one filter paired with a null detection in the other.

\end{enumerate}

Markov chains were run for 100 affine-invariant \textquotedblleft walkers\textquotedblright{} over 1000 initial burn-in iterations each to ensure adequate sampling of the parameter space given autocorrelation lengths of $<$50 iterations in all cases (finding that convergence was achieved within 10 autocorrelation lengths in all cases), and the results were then measured from 500 additional iterations.  In Table \ref{hbpartab}, we list, for each target field, the total number of stars in the CMD region used to build the LF $N_{obs}$, followed by the best fit (50th percentile) RC and RGBB parameters and their 1$\sigma$ confidence limits, corresponding to the 16th and 84th percentile; we do \textit{not} assume parameter distributions to be Gaussian.  For simplicity, we use $V(RC)$ to represent the ACS/WFC $F606W_{0}$ magnitude and $I(RC)$ to represent the ACS/WFC $F814W_{0}$ magnitude of the RC.  The observed LFs (binned only to permit a visual comparison) and the best fitting function are shown in Fig.~\ref{lfobsfig}, for both $F606W_{0}$ and $F814W_{0}$, and plots showing the posterior distributions for the LF parameters in each field are given in the Appendix. 

\begin{longrotatetable}
\begin{deluxetable}{lccccccccccc}
\tablecaption{Red Clump Parameters from Maximum Likelihood Fits \label{hbpartab}}
\tablehead{
\colhead{Field} & \colhead{$N_{obs}$} &\colhead{$I(RC)$} & \colhead{$\sigma$$I(RC)$} & \colhead{$V(RC)$} & \colhead{$\sigma$$V(RC)$} & \colhead{$V(RC)-I(RC)$} & \colhead{$EW_{RC}$} & \colhead{$\frac{N(RC)}{arcmin^2}$} & \colhead{$I^{RC}_{RGBB}$} & \colhead{$EW_{RGBB}$} & \colhead{$\sigma$$I(RGBB)$}
}
\startdata
Brown\_stream & 3250 & 24.345$^{+0.004}_{-0.005}$ & 0.108$^{+0.008}_{-0.008}$ & 25.088$^{+0.006}_{-0.006}$ & 0.130$^{+0.006}_{-0.007}$ & 0.743$^{+0.008}_{-0.007}$ & 1.367$^{+0.098}_{-0.100}$ & 109$^{+4}_{-5}$ & -0.466$^{+0.033}_{-0.028}$ & 0.220$^{+0.040}_{-0.033}$ & 0.129$^{+0.025}_{-0.021}$ \\
AMIGA4 & 900 & 24.542$^{+0.013}_{-0.014}$ & 0.130$^{+0.015}_{-0.013}$ & 25.289$^{+0.012}_{-0.013}$ & 0.134$^{+0.016}_{-0.015}$ & 0.746$^{+0.018}_{-0.016}$ & 1.36$^{+0.22}_{-0.20}$ & 17.1$^{+2.0}_{-2.1}$ & -0.406$^{+0.200}_{-0.089}$ & 0.39$^{+0.16}_{-0.15}$ & 0.229$^{+0.095}_{-0.093}$ \\
AMIGA6 & 139 & 24.494$^{+0.058}_{-0.069}$ & 0.257$^{+0.095}_{-0.072}$ & 25.209$^{+0.056}_{-0.062}$ & 0.235$^{+0.074}_{-0.057}$ & 0.713$^{+0.081}_{-0.077}$ & 1.35$^{+0.61}_{-0.45}$ & 2.7$^{+0.6}_{-0.7}$ & -0.15$^{+0.16}_{-0.30}$ & 0.25$^{+0.28}_{-0.18}$ & 0.24$^{+0.16}_{-0.14}$ \\
\enddata
\tablecomments{$V(RC)$ refers to the RC magnitude in $F606W_{0}$ and $I(RC)$ refers to the RC magnitude in $F814W_{0}$ in the ACS/WFC Vegamag system.  The values given for $EW_{RC}$, $EW_{RGBB}$, $\sigma$$I(RGBB)$, the density of RC stars $\frac{N(RC)}{arcmin^2}$, and the magnitude difference between the RC and RGBB $I^{RC}_{RGBB}$ are the $F814W_{0}$ values.}
\end{deluxetable}
\end{longrotatetable}

The remaining three metrics serve primarily as tests for differences in metallicity across our target fields, since such differences could systematically impact our use of the RC as a distance indicator. 

\subsection{Peak Color of the Red Clump \label{rccolorsect}}

\begin{figure}
\gridline{\fig{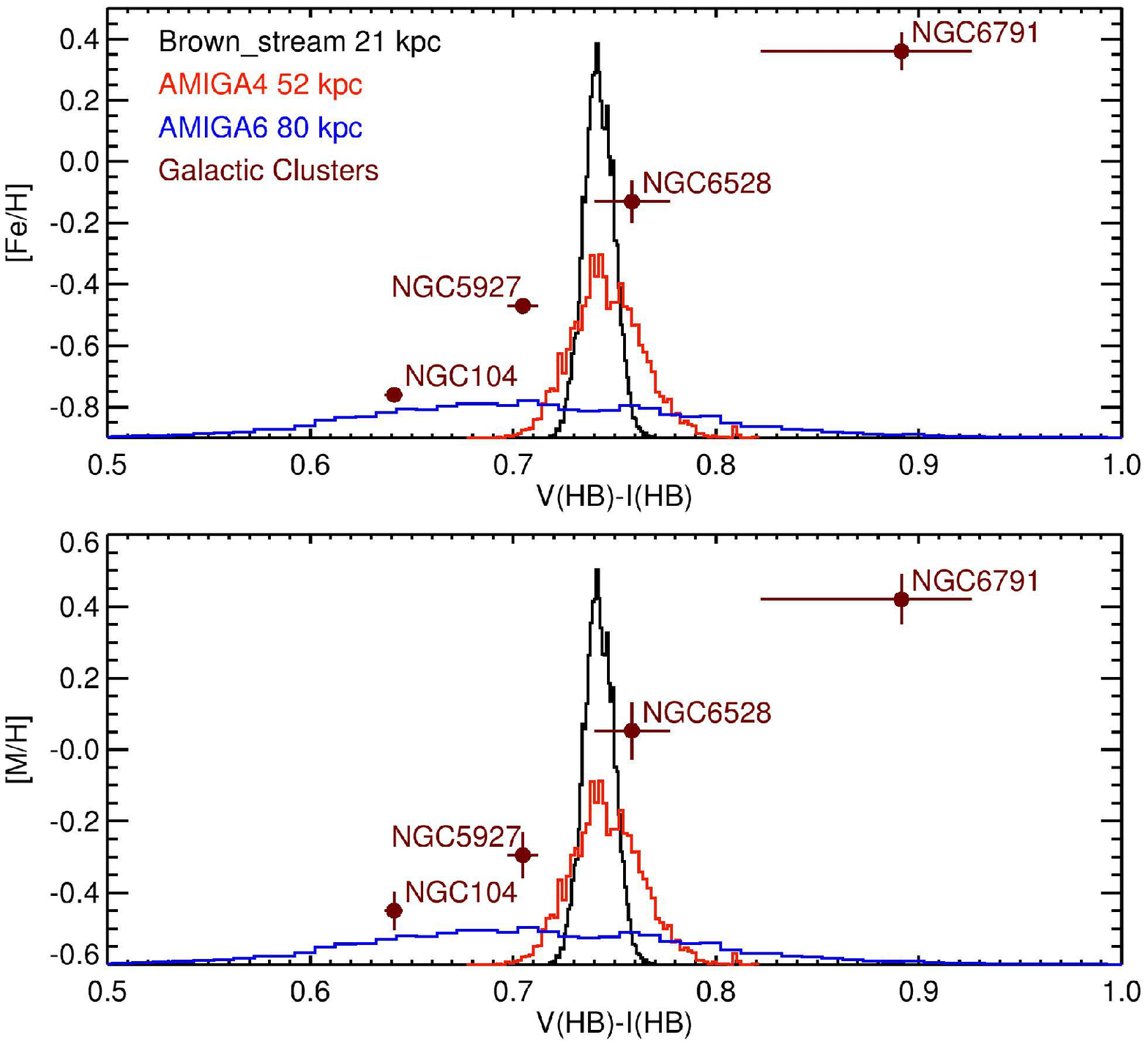}{0.49\textwidth}{}
	  \fig{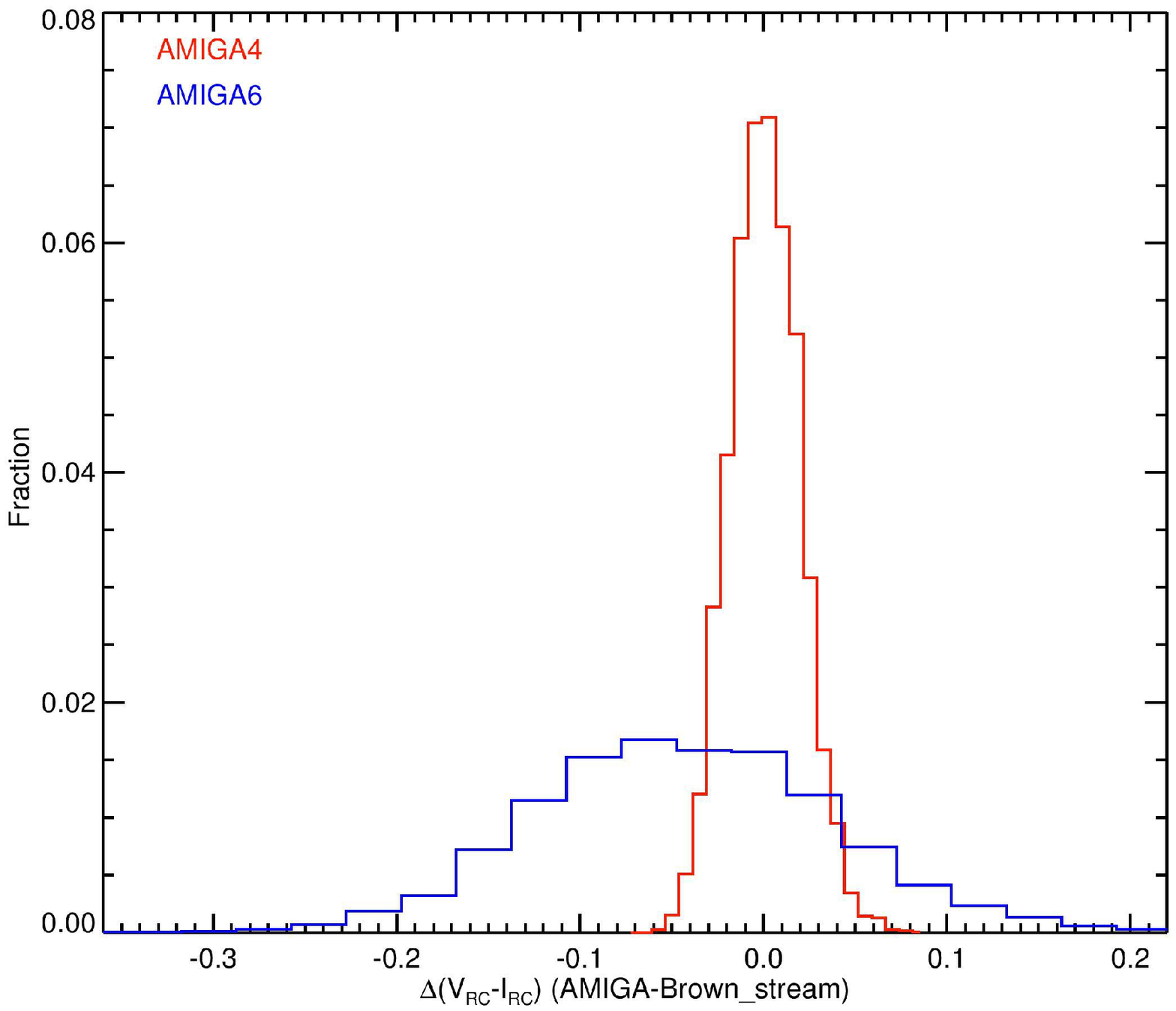}{0.49\textwidth}{}}
\caption{\textbf{Left: } Our empirical calibration of the RC peak color $V(RC)-I(RC)$ versus $[\rm{Fe/H}]$ (top) and global metallicity $[\rm{M/H}]$ (bottom), using metal-rich Galactic globular and open clusters as listed in Table \ref{ggctab}.  Histograms of the values we obtain for each of our three target fields are overplotted for comparison, color coded by field as in Fig.~\ref{lfobsfig} and normalized to the same total area.  \textbf{Right:} The RC peak color $V(RC)-I(RC)$ now compared \textit{differentially} with values seen for the 21 kpc Brown\_stream field over all MCMC iterations.  The broader color distribution for the AMIGA6 field is due to poor number statistics, although the RC is \textit{detected} in this field at a confidence level of 3$\sigma$ (see Table \ref{hbpartab}). \label{ggcfig}}
\end{figure}

To quantify the RC peak color, we use the difference of the peak magnitudes in the two filters rather than a mean or median taken over an entire color distribution, which has the advantage of tracing the dominant (i.e.~metal-rich GSS) population. 
Using the RC peak color as a metallicity diagnostic has the advantage that it is 
quite insensitive to age for ages over a few Gyr, demonstrated by both observational and theoretical evidence.  On the observational side, \citet{rc_age} find that the gradient of RC absolute magnitude with age differs by only their uncertainties of 0.003 mag/Gyr across the $Vri$ passbands using an asteroseismically selected sample (albeit containing only a small fraction of stars older than 8 Gyr).  Similarly, evolutionary models predict that the RC $(V-I)$ color at fixed metallicity varies by only $\sim$0.03 mag for ages beyond 3 Gyr, especially at high (near solar) metallicities \citep{gs01,girardirc}.  Therefore, we may leverage the small extinction (and correspondingly small extinction uncertainties) and measurements of the RC color to place stringent constraints on the metallicity gradient across our target fields based on their RC peak colors.  To derive an empirical relationship between RC peak color in $(F606W-F814W)_{0}$ versus both $[\rm{Fe/H}]$ and global metallicity $[M/H]$ \citep{feh2mh}, we use publicly available photometry of the metal-rich Galactic globular clusters NGC 104 (47 Tuc) and NGC 5927 from \citet{ataggc} and supplement them with additional archival imaging of the super-solar-metallicity old open cluster NGC 6791 and the metal-rich bulge globular cluster NGC 6528 (\citealt{lagioia6528}, Cohen et al.~in prep.).  We perform maximum likelihood fits as described in Sect.~\ref{rcmagsect}, assuming recent literature distance, reddening, $[\rm{Fe/H}]$ and $[\alpha\rm{/Fe}]$ values as listed in Table \ref{ggctab} and restricting the fits to stars with -1.1$\leq$$M_{814W}$$\leq$0.9.  The resulting trends of $V(RC)-I(RC)$ versus $[\rm{Fe/H}]$ and $[\rm{M/H}]$ are shown in Fig.~\ref{ggcfig}, and imply that $V(RC)-I(RC)$ has a metallicity sensitivity of $\sim$0.3 mag per dex.   For comparison, the histograms of the MCMC results from our three target fields are overplotted, color coded by field as in Fig.~\ref{lfobsfig}.  It is apparent that the GSS fields at 21, 52 and 80 kpc have RC colors which agree to within their uncertainties (albeit with large uncertainties for the 80 kpc field owing to poor number statistics), and the peak RC color we measure for the 21 and 52 kpc fields constrains their metallicities to be only slightly subsolar, in agreement with the metallicities we measure photometrically from the upper RGB in Sect.~\ref{photmetsect}.  In the right panel of Fig~\ref{ggcfig}, we compare RC peak color variations across our target fields in a strictly differential sense, where the difference between the RC peak color of the AMIGA4 and AMIGA6 fields is shown compared to the Brown\_stream field over all MCMC iterations.  Given our empirically derived metallicity gradient above, the RC peak color difference between the 21 kpc and 52 kpc fields places a 1$\sigma$ constraint of $\sim$0.15 dex on their difference in $[\rm{Fe/H}]$.  Meanwhile, the broader constraints on the AMIGA6 field easily allow such a difference, and give a 1$\sigma$ limit of $\sim$0.5 dex on the metallicity decrease seen from 21 to 80 kpc.

\begin{deluxetable}{lccc}
\tablecaption{Parameters of Galactic Clusters \label{ggctab}}
\tablehead{
\colhead{Cluster} & \colhead{$E(B-V)$} & \colhead{$[\rm{Fe/H}]$} & \colhead{$[\alpha\rm{/Fe}]$}
}
\startdata
NGC 104 & 0.032 (1) & -0.76$\pm$0.02 (2) & 0.42$\pm$0.06 (3) \\
NGC 5927 & 0.415 (1) & -0.47$\pm$0.02 (4) & 0.25$\pm$0.08 (4) \\
NGC 6528 & 0.54 (5) & -0.13$\pm$0.07 (6) & 0.26$\pm$0.05 (6) \\
NGC 6791 & 0.14 (7) & 0.36$\pm$0.06 (7) & 0.09$\pm$0.01 (7) \\
\enddata
\tablecomments{Literature source for each quantity is given in parenthesis, the corresponding references are as follows: (1) \citet{vandenbergages}; (2) \citet{c09}; (3) \citet{c10}; (4) \citet{aldo5927}; (5) \citet{calamida6528}; (6) \citet{bruno6528}; (7) \citet{cunha6791}.}
\end{deluxetable}

\subsection{Horizontal Branch Color Distribution \label{hbcolsect}}

The color ratio of blue to total horizontal branch stars may also be useful to constrain age and/or metallicity differences between our target fields.  To isolate the HB in the CMD, we use the selection box shown in Fig.~\ref{cmdfig}, built using the well-populated Brown\_stream field as a template.  We then measure the ratio of HB stars in the box with $(F606W-F814W)_{0}$$<$0.2 and 0.4 to all stars in the box, 
accounting for foreground contaminants, background galaxies 
and their uncertainties (see Sect.~\ref{contamsect}), propagated in quadrature.  For the Brown\_stream field, we subtract 23\% contamination from the Brown\_halo11 field \citep{jason_gss_spec} by counting the number of stars in the entire HB selection region as well as those with $(F606W-F814W)_{0}$$<$0.2 and 0.4 for the Brown\_stream field and then normalizing these counts to the total number of observed stars in each region in the Brown\_halo11 field.  We find that the AMIGA4 field at 52 kpc has a redder HB morphology than the Brown\_stream field at more than 1$\sigma$: The fraction of HB stars with $(F606W-F814W)_{0}$$<$0.2 is 0.012$\pm$0.007 at 52 kpc, compared to 0.027$\pm$0.004 at 21 kpc, and similarly the fraction of HB stars with $(F606W-F814W)_{0}$$<$0.4 is unchanged in the 52 kpc field (0.012$\pm$0.008), compared to 0.039$\pm$0.005 at 21 kpc.  For the 80 kpc field, the fraction of blue HB stars is statistically equivalent to zero after accounting for background galaxy contamination, but small number statistics result in larger uncertainties.  This metric could, in principle, be biased due to the width in magnitude of the HB selection box, since more metal-rich populations would have their RGBB shifted faintward of the selection box boundary on the RGBB.  However, the similarity between the RGBB properties of the fields at 21 and 52 kpc (see Table \ref{hbpartab}) implies that the difference in their HB morphology is not affected by the RGBB.

An alternative way to quantify the HB color was employed by \citet{martindolphotcuts}, who use red and blue HB boxes which are fixed in color but allowed to move in magnitude in order to be centered on the RC.  They use $\eta$ to denote the fraction of stars located in the blue box compared to the total in both boxes.  If we use their color boundaries for the blue and red HB boxes, 
we find a fraction of $\eta$=0.060$\pm$0.007 at 21 kpc, compared to $\eta$=0.034$\pm$0.012 at 52 kpc, indicating a redder HB morphology for the 52 kpc field, consistent with the above analysis.  The values of $\eta$ we find are consistent with the more luminous dwarf spheroidal (dSph) galaxies analyzed in \citet{martindolphotcuts}, most of which lie in their \textquotedblleft Off-plane\textquotedblright{} sample.  While this result is interesting at face value, the authors point out that the $\eta$ metric is biased.  Most relevantly to our results, this bias operates such that the red cutoff of their red HB box, at $(F606W-F814W)_{0}$=0.75, will systematically yield lower values of $\eta$ for clusters with near-solar (or higher) $[\rm{Fe/H}]$ because as metallicity increases, the RGB is shifted increasingly redward of the red HB box.
While this appears to cause little intra-cluster bias among their sample, the near-solar metallicity of the 52 kpc field could inflate $\eta$ since a higher fraction of RGB stars fall redward of the red HB box.

\subsection{Photometric Metallicity Distribution \label{photmetsect}}

\begin{figure}
\gridline{\fig{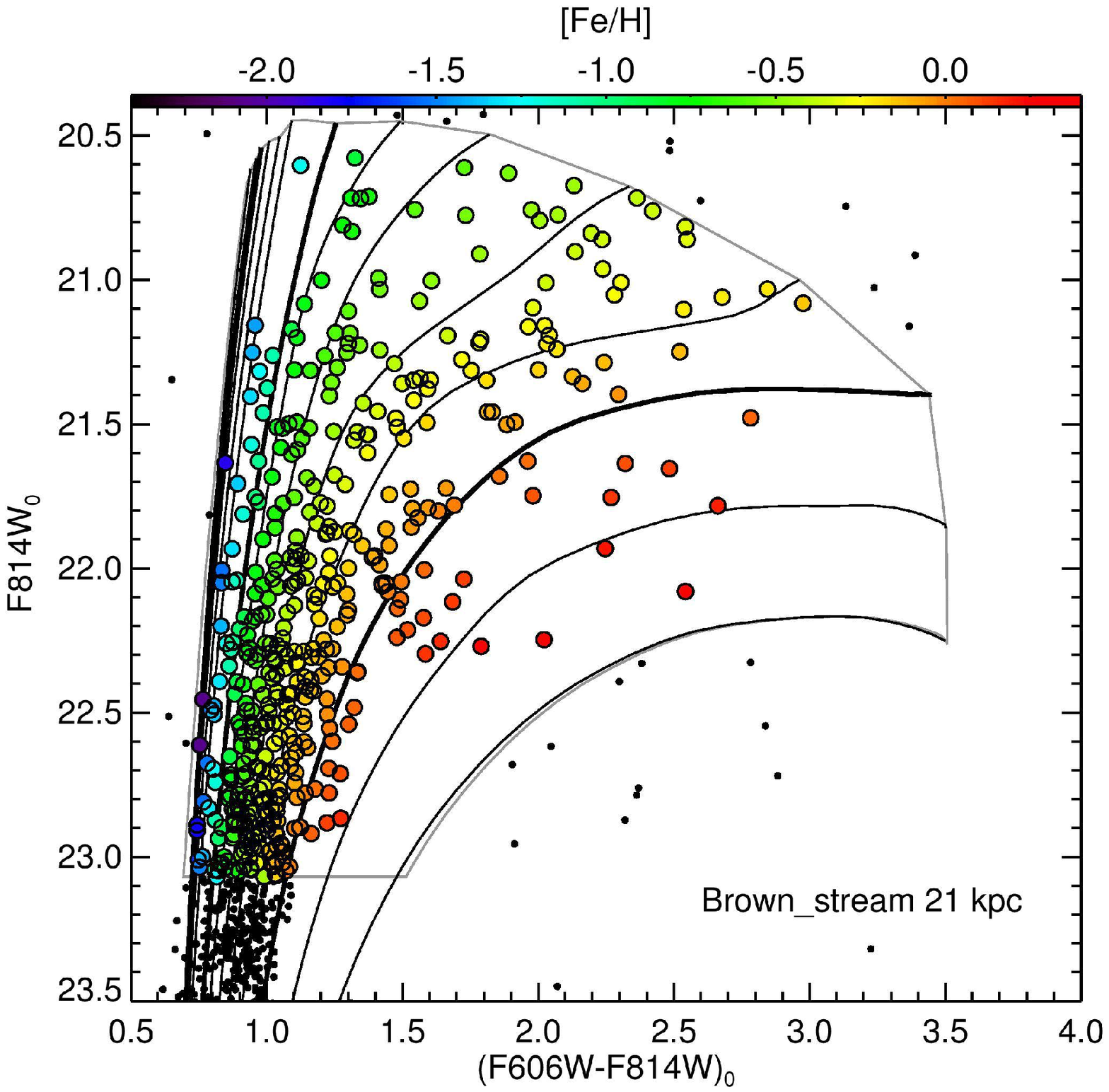}{0.33\textwidth}{}
	  \fig{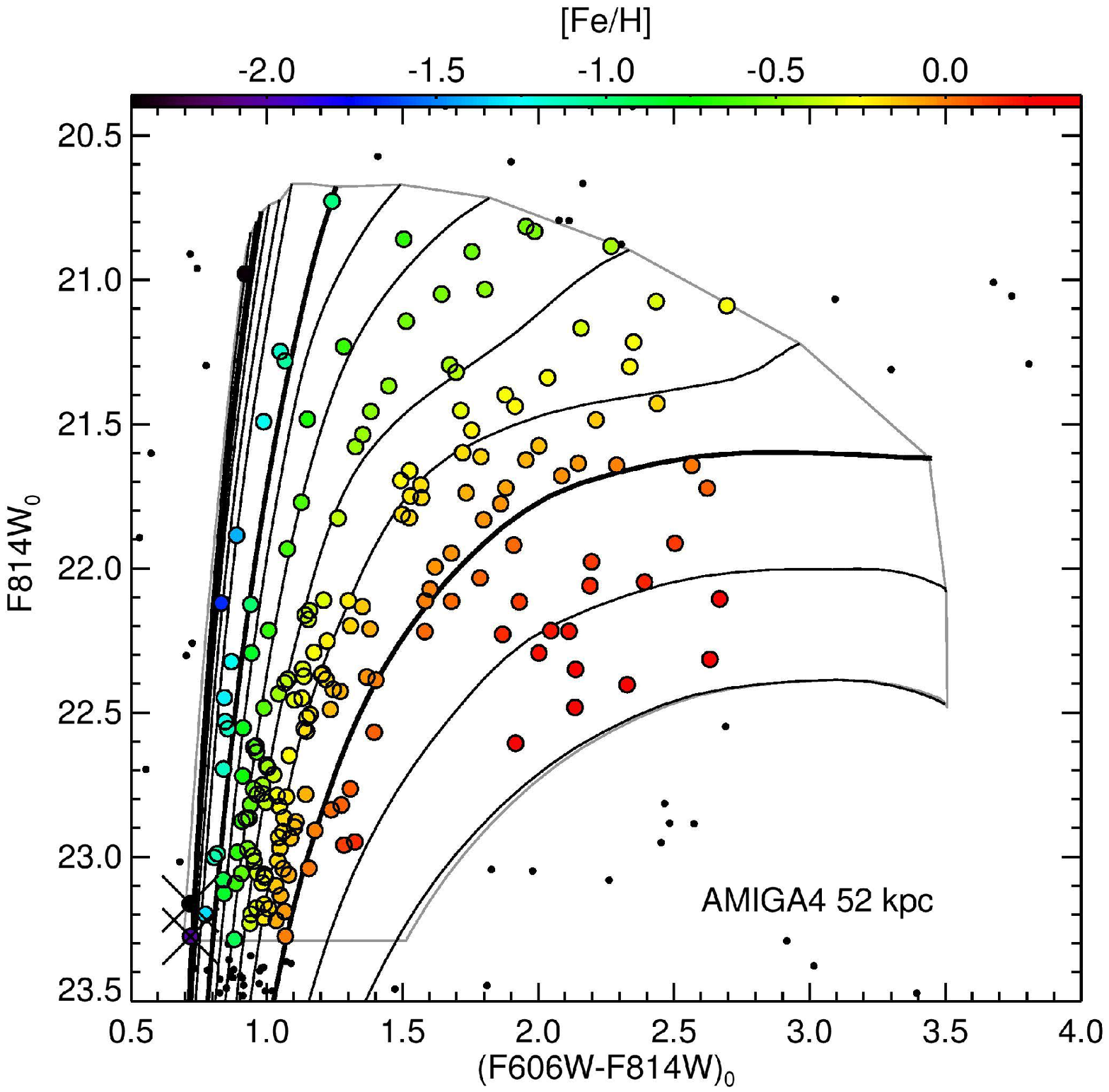}{0.33\textwidth}{}
	  \fig{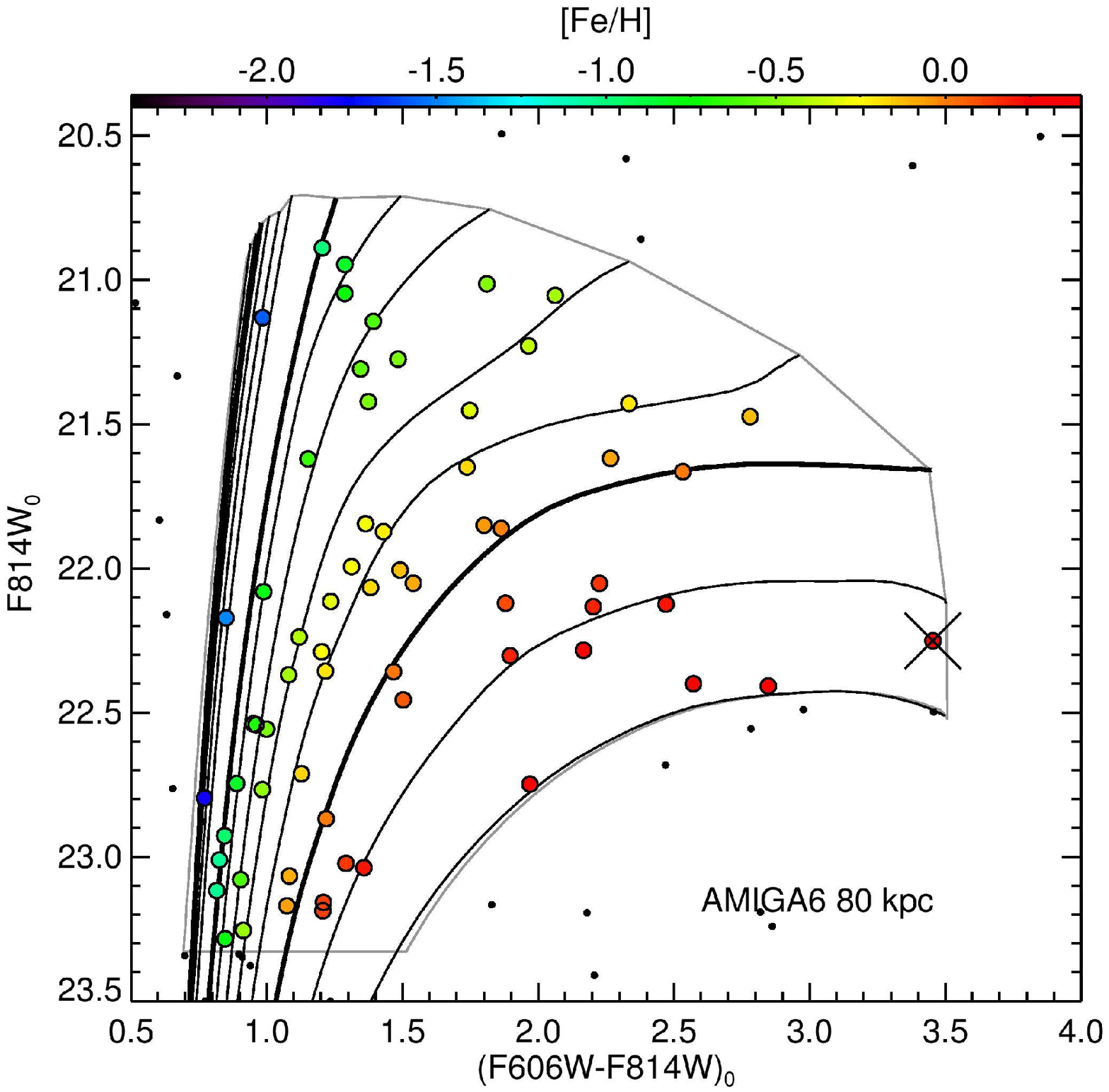}{0.33\textwidth}{}}
\caption{CMDs of the three GSS fields at projected distances of (left to right) 21, 52 and 80 kpc from Andromeda, illustrating photometric metallicities derived by interpolating in an isochrone grid.  In each panel, the region used to select M31 RGB stars, defined using Victoria-Regina isochrones with -2.4$\leq$$[\rm{Fe/H}]$$\leq$+0.4 plus 1$\sigma$ photometric errors, is outlined in grey.  Individual isochrones are shown as curved black lines, in increments of 0.2 dex from right to left, with isochrones corresponding to $[\rm{Fe/H}]$=[0,-1,-2] shown as thick lines.  Stars inside the grey selection region and more than 1.3 mag brightward of the HB, used to construct the photometric metallicity distribution, are color-coded by metallicity according to the colorbar at the top of each plot.  Stars falling inside the selection region but beyond the isochrones, with metallicities derived via interpolation outside the isochrone grid, are overplotted as crosses, although they constitute $\leq$1\% of the sample in all cases. \label{photmetcmdfig}}
\end{figure}

\begin{figure}
\plottwo{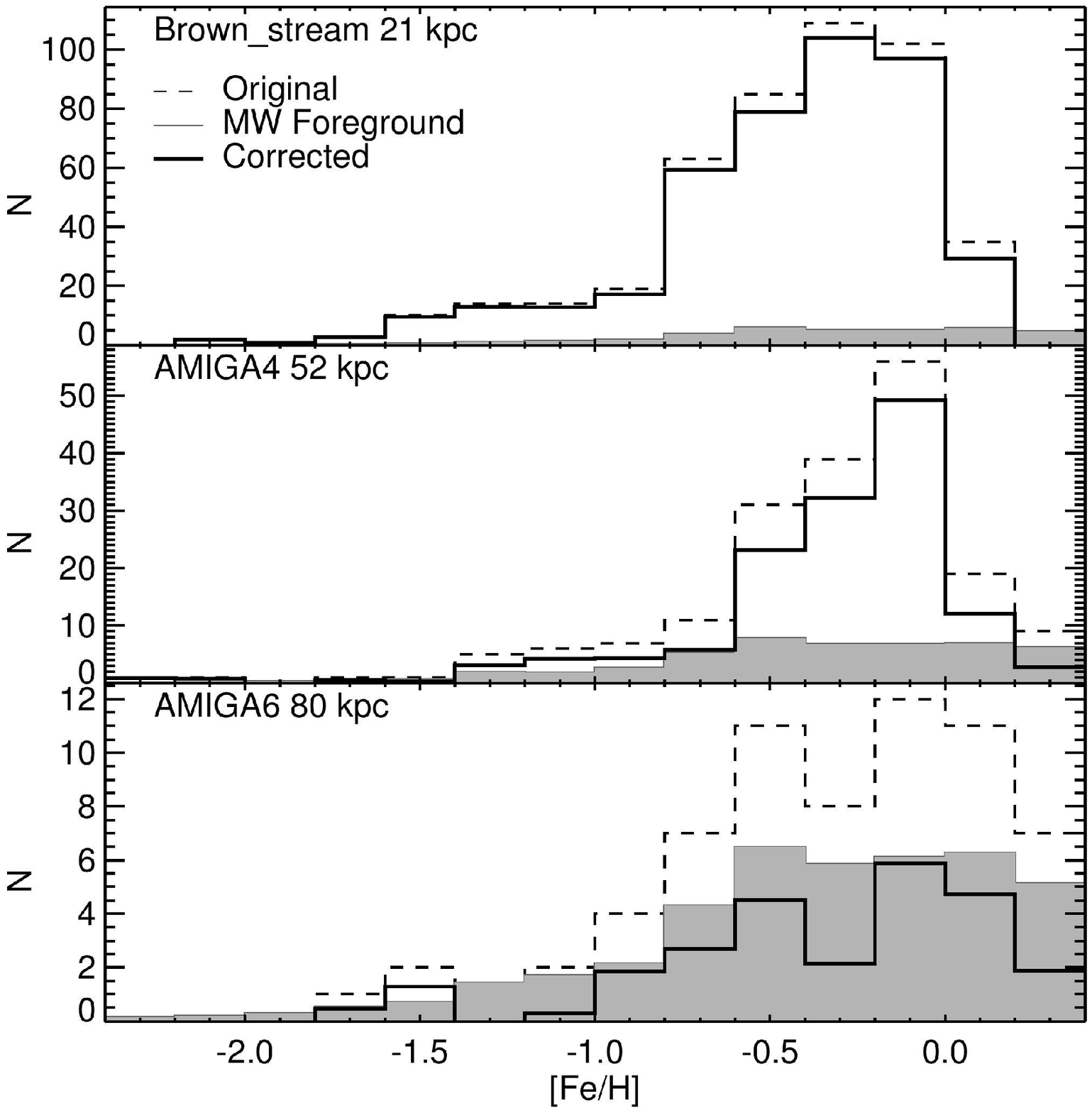}{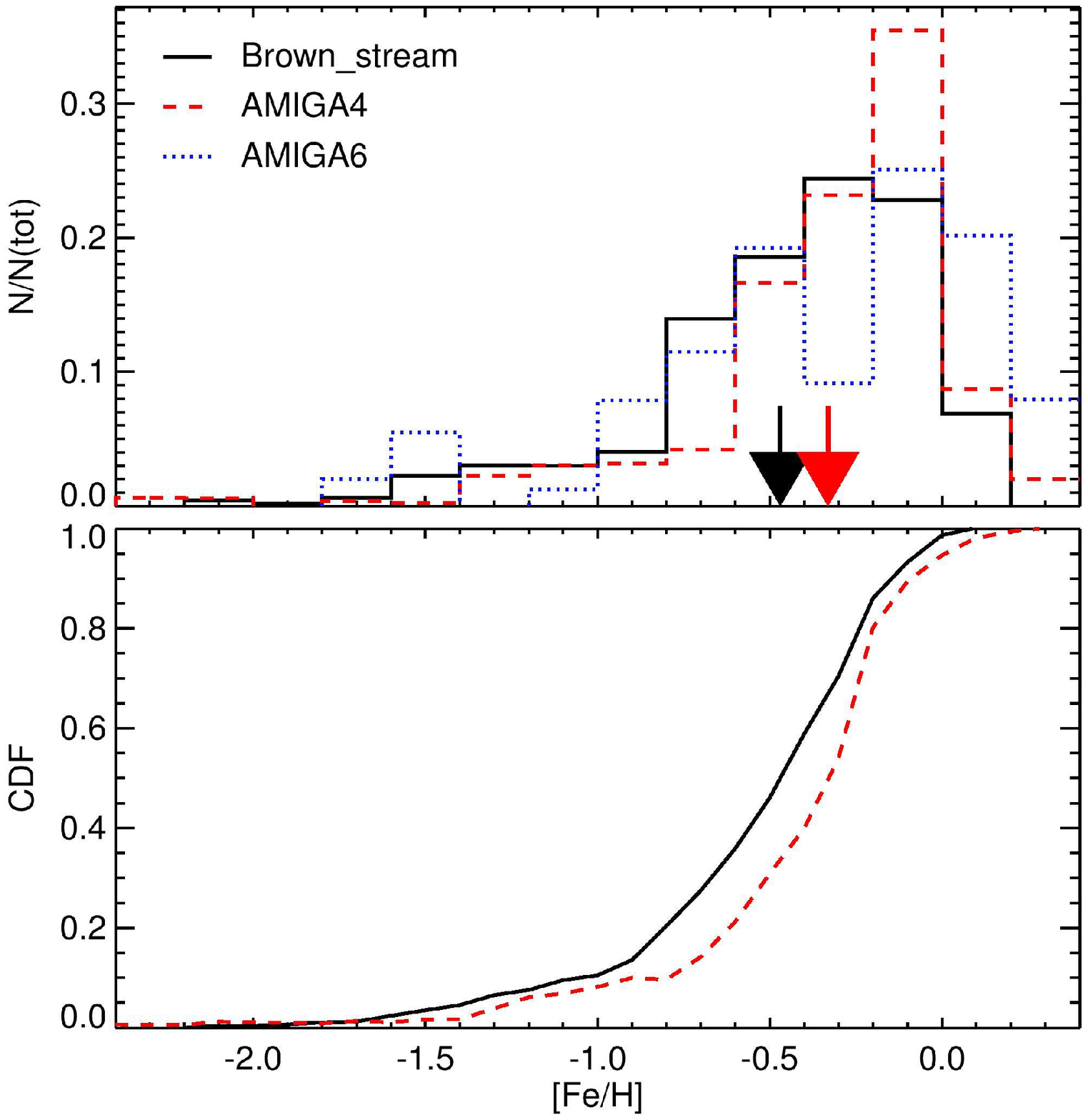}
\caption{\textbf{Left:} Photometric metallicity distributions for three GSS fields at projected radii of (top to bottom) 21, 52 and 80 kpc from M31.  The raw distribution is shown by the dashed line, before being decontaminated using the area-scaled TRILEGAL Galaxy model (shaded region) to produce the Milky-Way-decontaminated metallicity distribution (thick solid line). \textbf{Right:} Differential (top) and cumulative (bottom) photometric metallicity distributions for the target fields plotted together.  The arrows in the top right plot indicate the median photometric metallicity in the 21 kpc Brown\_stream field (black) and the 52 kpc AMIGA4 field (red).  The AMIGA6 field is excluded from the bottom right plot because the Galaxy model gives an oversubtraction of the foreground metallicity distribution in some bins, and its low stellar density renders its photometric $[Fe/H]$ extremely sensitive to the assumed foreground Galaxy model (see text). \label{photmethists}}
\end{figure}

In addition to using the RC and HB colors as a diagnostic of metallicity differences between our fields, we calculate metallicity distributions photometrically by interpolating in an isochrone grid as a check on 
 the constancy of the metallicity along the GSS from 21 kpc to 52 kpc. 
We use the same CMD selection region as described in Sect.~\ref{rcmagsect}, which has been shifted to account for differences in line of sight distance among our target fields, and impose the additional restriction that stars must lie $\geq$1.3 mag brightward of the RC in an attempt to minimize contamination from the AGB bump ($\Delta$$I (HB-AGBB) \sim$1.06; \citealt{natafagbb}).  
The $[\rm{Fe/H}]$ value for each star is determined by linear interpolation among the isochrones, and in Fig.~\ref{photmetcmdfig} we show CMDs for each field, where stars are color-coded by their metallicity according to the isochrones.  Since the boundaries of the CMD selection region have been broadened in color by the 1$\sigma$ photometric errors, there is a small fraction of stars lying outside the isochrones but inside the selection region, but these stars (shown using crosses in Fig.~\ref{photmetcmdfig}) constitute $\leq$1\% of the sample in all cases and affect the median $[\rm{Fe/H}]$ at the 0.02 dex level.
Histograms of the photometric $[\rm{Fe/H}]$ distribution in each of our three target fields are shown both individually and together in Fig.~\ref{photmethists}.  In the left panels of Fig.~\ref{photmethists}, the photometric metallicity distributions for each of the three target fields are shown individually, where the raw distribution of all sources falling in the selection region is shown as a dashed line.  We use the TRILEGAL Galaxy model \citep{trilegal1,trilegal2} to decontaminate the foreground Milky Way contribution (see Sect.~\ref{mwcontamsect}), shown as a grey shaded region, and the resulting foreground-subtracted metallicity distributions are shown as thick solid lines. We obtain predicted foreground contamination fractions of 8\% and 25\% for the 21 and 52 kpc fields respectively, while our background galaxy contamination model (see Sect.~\ref{galcontamsect}) predicts 0$\pm$3 background galaxy contaminants for all three fields over the upper RGB region, statistically equiavalent to zero.   

A check on the TRILEGAL predictions is provided by 
 the spectroscopic results of \citet{gilbert_gss,gilbert_sb_global}, and to perform a direct comparison between the two, we shift the faint magnitude limit of the upper RGB region brightward by 0.6 mag in order to sample the same CMD region from which their spectroscopic targets were selected.  Counting the number of secure Milky Way dwarf contaminants and the number of secure M31 RGB stars reported in table 1 of \citet{gilbert_sb_global}, their results imply a foreground contamination fraction of 15\% compared to 11\% from the TRILEGAL simulations in the H13s SPLASH field, which is colocated with the 21 kpc Brown\_stream field.  For their fields d1 and a13, which bracket the spatial location of our AMIGA4 field (see Fig.~\ref{posfig}), they find contamination fractions of 48\% and 57\% respectively, compared to 33\% predicted by the TRILEGAL model.  While this appears to indicate that the TRILEGAL model underpredicts foreground contamination by over 40\% in the latter case, the size of the spectroscopic sample is fairly small ($<$100 stars each in the d1 and a13 fields) and has a selection function subject to various systematics discussed in \citet{gilbert_method} and \citet[][see their sect.~3.2]{gilbert_sb_global}.  In any case, the similarity of the foreground and GSS metallicity distributions renders the foreground-subtracted median metallicity in our target fields fairly insensitive to changes in the assumed foreground contamination fraction.  For example, even doubling the foreground contribution from the TRILEGAL models only impacts our median photometric metallicities at $\leq$0.02 dex in the 21 and 52 kpc fields.

In the right panel of Fig.~\ref{photmethists}, we plot the differential and cumulative metallicity distributions of the 21 and 52 kpc fields together, where the arrows indicate the median $[\rm{Fe/H}]$ for each field.  
We find that the 52 kpc field is actually the most metal-rich, with a median $[\rm{Fe/H}]$=
-0.33$\pm$0.04  
compared to 
-0.47$\pm$0.02 in the 21 kpc field (uncertainties represent the quadrature sum of contributions from photometric errors and stochasticity in the target and foreground populations, assessed via bootstrap resampling with repetition).  
Turning to the width of the $[\rm{Fe/H}]$ distributions, calculated as the interval between the 16th and 84th percentiles (since they are clearly non-Gaussian) and denoted as 
$\Delta$$[\rm{Fe/H}]$,
the 52 kpc field has a somewhat more narrow $[\rm{Fe/H}]$ \textit{distribution}, with a 16th to 84th percentile width of $\Delta$$[\rm{Fe/H}]$=0.52$^{+0.08}_{-0.07}$ dex compared to 
0.65$\pm$0.03 dex for the 21 kpc field, consistent with the fairly extended star formation histories measured in that field by \citet{brownsfh} and \citet{bernard_sfh}.  

The photometric $[\rm{Fe/H}]$ distribution of the 80 kpc field has been excluded from the right panel of Fig.~\ref{photmethists}, since the stellar density there is so low that its photometric metallicity is extremely sensitive to assumptions on the foreground Milky Way population: The TRILEGAL model predicts a contamination fraction of 64\%,
and slightly oversubtracts some of the $[\rm{Fe/H}]$ bins as seen in the left panel (although at a statistically insignificant level).

An additional source of systematic uncertainty in photometric metallicities measured on the upper RGB is due to contamination from the AGB, the details of which depend on age, metallicity, $\alpha$-enhancement, and mass loss, among others.  However, to get a rough estimate of the effect of AGB stars on our photometric metallicity distributions, we use synthetic CMDs generated using the BaSTI database \citep{basti1,basti2,basti3} to compare RGB-only distributions to those including AGB stars.  By varying the input $[\rm{Fe/H}]$ (between -1 and 0), age (from 6-12 Gyr), and mass loss exponent (from 0.2 to 0.4), the shift in median $[\rm{Fe/H}]$ is consistently close to -0.1 dex, and AGB contaminants act mainly to extend the low-metallicity tail of the metallicity distribution, broadening the 1$\sigma$ lower envelope by as much as $\sim$0.25 dex for a population with $[\rm{Fe/H}]$=-1.

In summary, all three of our photometric metallicity diagnostics provide consistent results.  The peak color of the RC (Sect.~\ref{rccolorsect}), the HB color distribution (Sect.~\ref{hbcolsect}) and photometric metallicities on the upper RGB (Sect.~\ref{photmetsect}) all imply a slightly ($\sim$0.15 dex) increasing metallicity moving from the 21 kpc Brown\_stream field to the 52 kpc AMIGA4 field.  For the 80 kpc AMIGA6 field, all three metallicity diagnostics also allow for either a similar metallicity as the 52 kpc field or a decrease by as much as $\sim$0.5 dex given large statistical uncertainties.

\subsection{Sources of Contamination \label{contamsect}}
\subsubsection{Milky Way Foreground \label{mwcontamsect}}

Previous studies of fields towards the GSS and M31 halo were able to make use
of an ensemble of photometric \textit{and} spectroscopic information in order
to construct samples cleaned of Milky Way foreground contaminants \citep[e.g.][]{gilbert_method,gilbert_sb_global}.  
In our case, we have only CMDs at our disposal to mitigate foreground contamination, so  
the locations of our CMD selection boxes in Fig.~\ref{cmdfig} were set to exclude Milky Way stars to the greatest extent possible.  
Foreground contamination as a function of color and magnitude is quantified
using the TRILEGAL Galaxy model with the appropriate ACS/WFC bandpasses and foreground Milky Way extinction.  By performing random draws from an area of 1 deg$^2$ ($\sim$300 times the ACS/WFC field of view) we may calculate the effect of Milky Way foreground contaminants and their stochastic variation on our results.
The density of Milky Way foreground contaminants predicted by the TRILEGAL models is illustrated in the upper left panel of Fig.~\ref{contamfig}, and a subset of model foreground stars corresponding to an area of 0.1 deg$^2$ is color coded according to their Galactic component.  For reference, the CMD selection region used to build our RC+RGB LFs (Sect.~\ref{rcmagsect}) is shown as a solid black solid line, while the region used to analyze the HB color distribution (Sect.~\ref{hbcolsect}) is shown as a dashed black line and the region used to obtain photometric metallicity distributions on the upper RGB (Sect.~\ref{photmetsect}) is shown as a dotted black line.

The TRILEGAL models predict 
contamination fractions of 2\% towards the AMIGA6 field and $<$1\% for the AMIGA4 and Brown\_stream fields for both the RC+RGB region and the HB region in the target field CMDs.  However, current Galaxy models have been known to underpredict foreground contamination due to their failure to account for recently discovered halo substructure \citep[e.g.][]{pandas_substructure,martin_pandas_substructure}.  In fact, our AMIGA fields are roughly coincident with the location of the TriAnd2 overdensity \citep{martin_triand2,triand_sdss}.  However, using the same models and distance moduli as \citet{martin_triand2,martin_pandas_substructure} and conservatively assuming 300 stars per deg$^2$ in the magnitude range 21.4$\leq$$i_{0}$$\leq$23.4 (see their figs.~1-2 and \citealt{triand_sdss}), the number of TriAnd2 main sequence stars predicted to intrude into our RC+RGB selection region is $\sim$0.2 per arcmin$^2$ assuming a lognormal stellar mass function.  Their results illustrate that towards our AMIGA fields, the contribution of TriAnd2 generally dominates over that of TriAnd, but even summing the two contributions assuming they are equal still yields $<$0.5 stars per arcmin$^2$ 
for a contamination fraction of $<$1\% ($<$6\%) towards the AMIGA4 (AMIGA6) fields.  Our RC analysis is
almost certainly unaffected by such a small fraction of contaminants for two reasons.  First, any TriAnd contaminants are distributed over $>$2 mag in the RC+RGB region (see Fig.~\ref{cmdfig}).  Second, TriAnd or TriAnd2 main sequence
stars are unlikely to extend faintward to overlap the GSS RC itself since TriAnd2 (the more distant of the two) has a small (heliocentric) distance distribution (D$_{\sun}$=22-32 kpc; \citealt{martin_pandas_substructure,triand_distance}).

\subsubsection{Background Galaxies \label{galcontamsect}}

We have set our \texttt{Dolphot} PSF photometry parameters and quality cuts to be optimized for rejection of background galaxies.
However, 
it is crucial that we quantify any remaining
contamination from background galaxies with apparently stellar PSFs lest their CMD loci bias any of the parameters we wish to measure.  Therefore, we have searched the \textit{HST} archive for 
relatively \textquotedblleft blank\textquotedblright{} high Galactic latitude ACS/WFC fields observed with identical filters and similar
photometric depth.
The blank fields we have selected are listed in Table \ref{blankfldtab}, and include the CT344 field similarly employed by \citet{ghosts} as well as a subset of the parallel imaging taken as part of the Hubble Frontier Fields program \citep{hff}.
 
By choosing very sparse fields at high Galactic latitudes (5/6 of the blank fields have $|$$B$$|$$>$60) and performing PSF photometry identically as described in Sect.~\ref{photsect}, the difference between the CMD loci of the recovered stars (passing our photometric quality cuts) and the TRILEGAL Galaxy model (i.e.~foreground) prediction towards each field yields the density of non-rejected background galaxies as a function of CMD location.  
While this procedure assumes that the Galaxy model correctly predicts the Milky Way population towards the selected high-Galactic-latitude fields, it is pessimistic in the sense that an underprediction of the
Milky Way halo population by the Galaxy model will yield an \textit{overestimate} of the background galaxy density.

\begin{figure}
\gridline{\fig{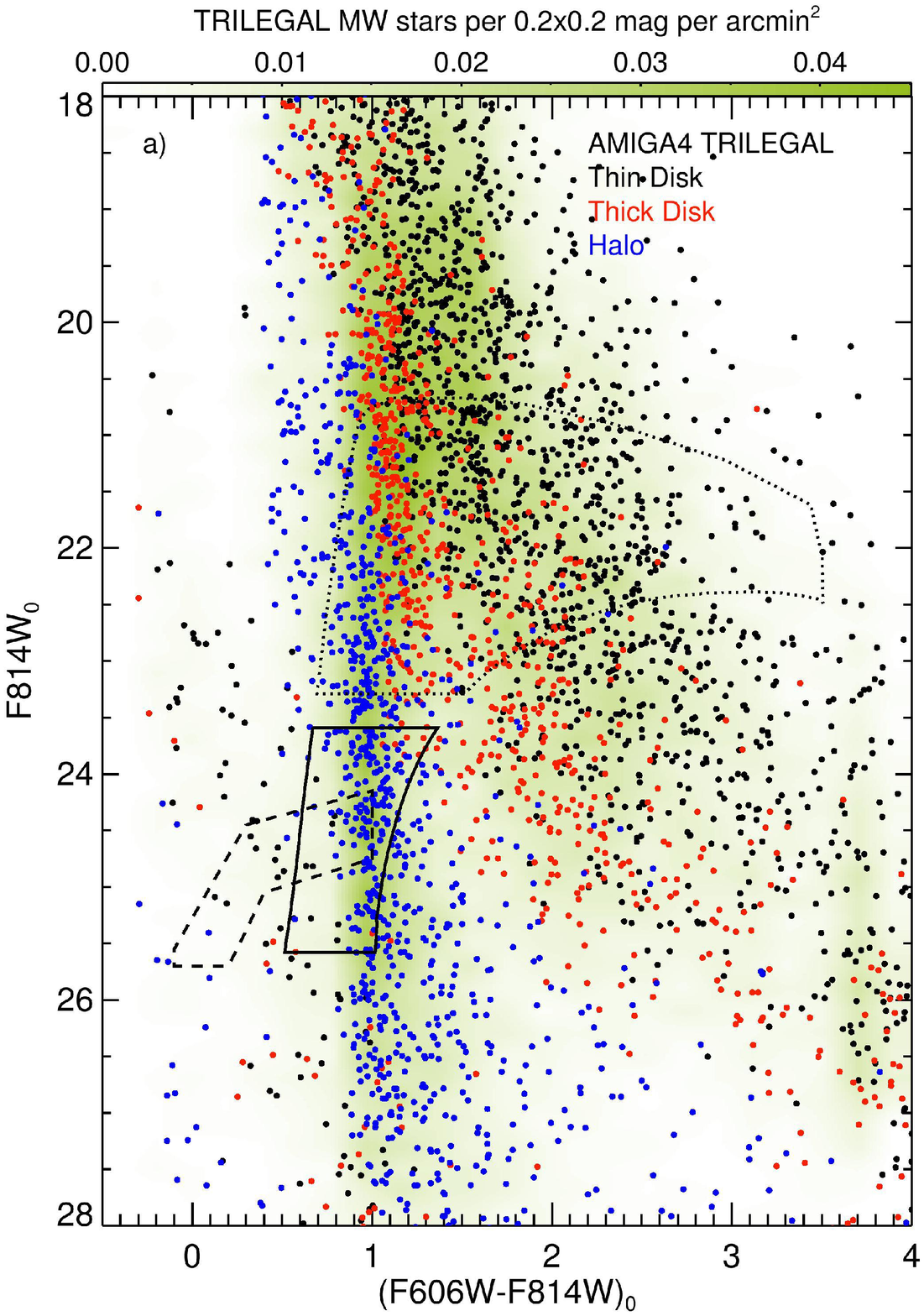}{0.33\textwidth}{}
	  \fig{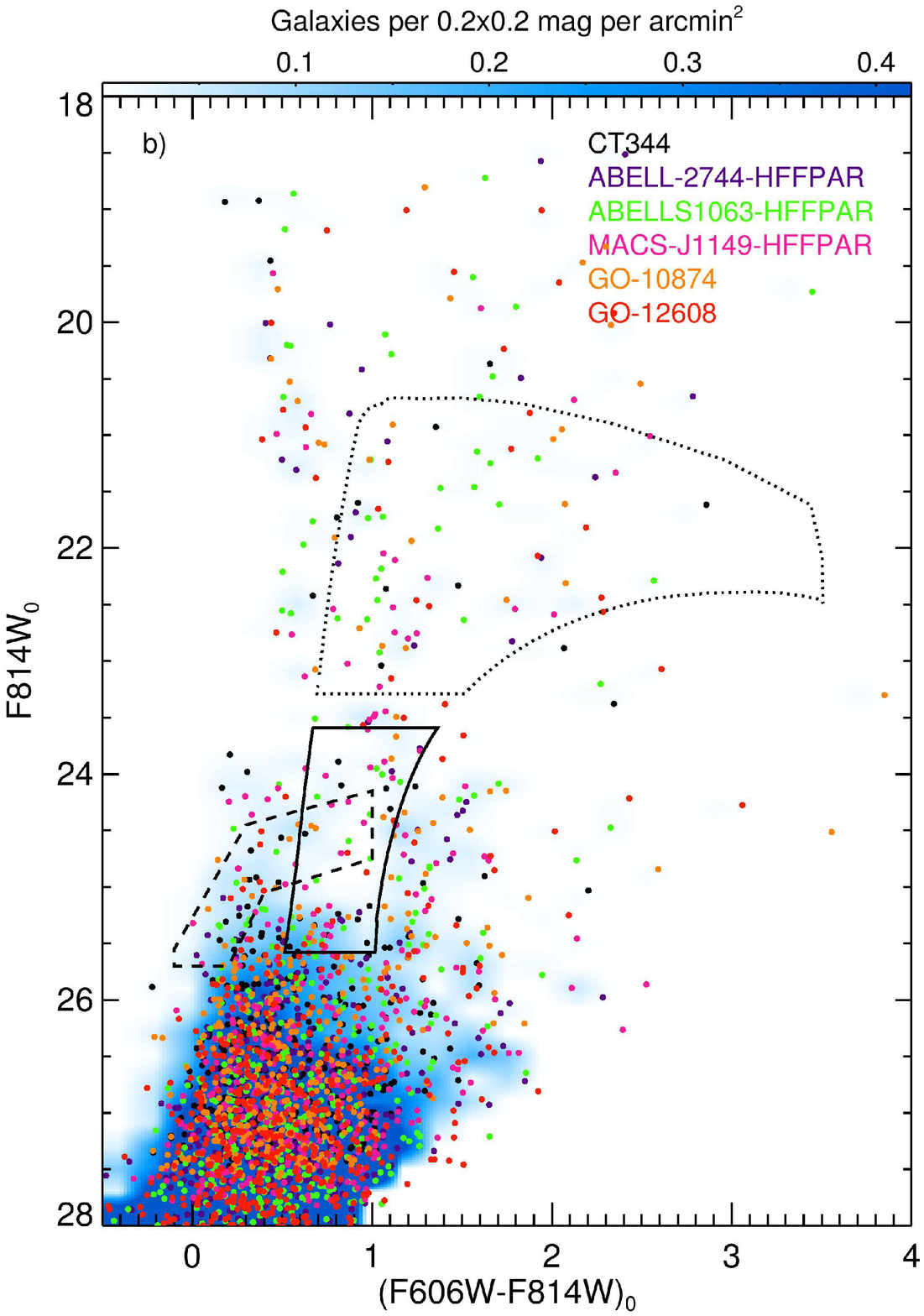}{0.33\textwidth}{}
	  \fig{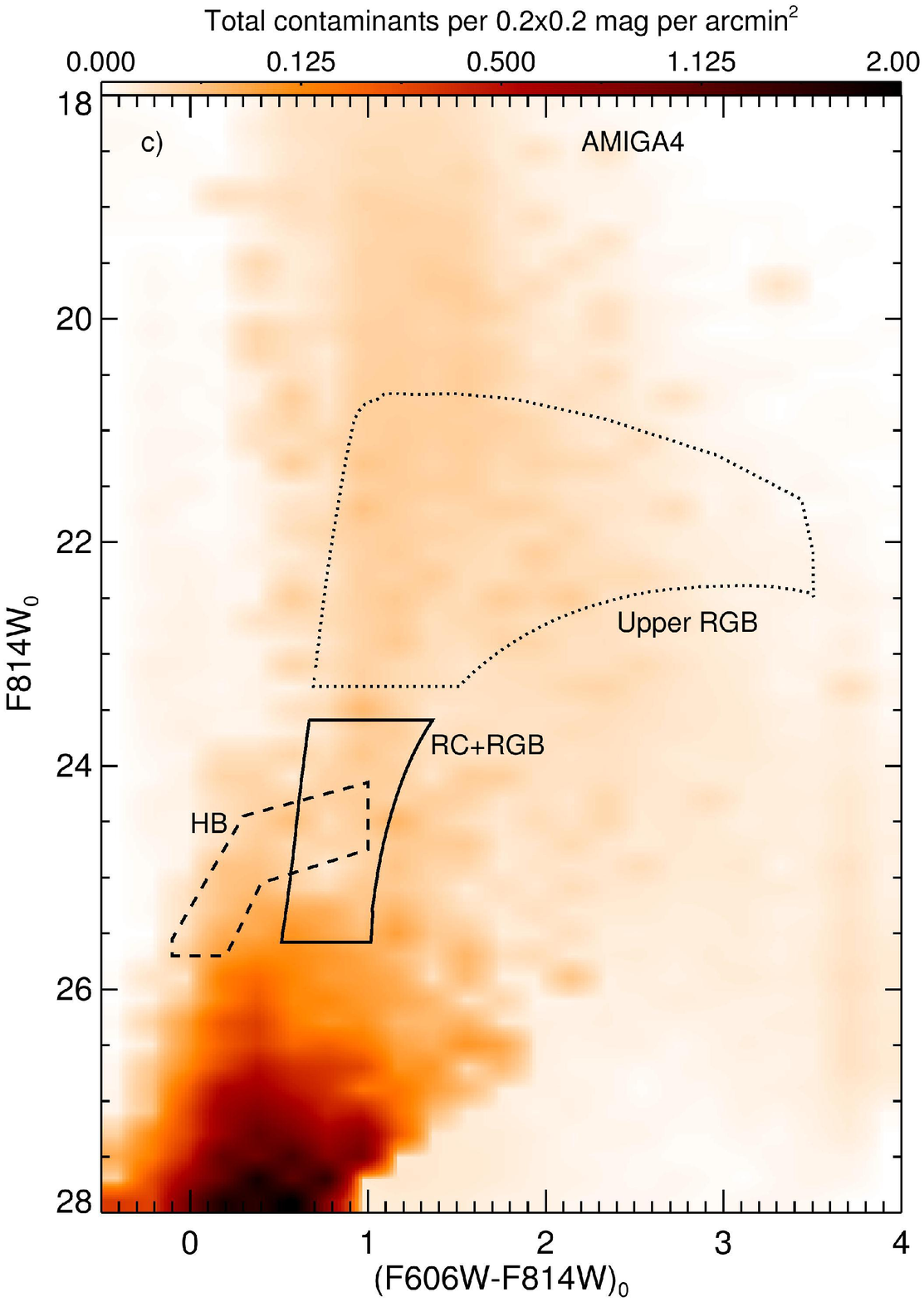}{0.33\textwidth}{}}
\caption{Model of foreground Milky Way plus background galaxy contaminants, for the example of the AMIGA4 field.  \textbf{(a):} Milky Way foreground contaminants towards the AMIGA4 field based on the TRILEGAL model.  Stars are color coded by their Galactic component (thin disk, thick disk, or halo) in the model.  The color bar at the top gives the number of contaminants per 0.2$\times$0.2 mag color-magnitude box per arcmin$^2$.  \textbf{(b):} Background galaxy contaminants from
six high-latitude \textquotedblleft blank\textquotedblright{} fields.  
All recovered sources are shown, color coded by field, and the shading indicates the completeness-corrected density of background galaxies obtained by subtracting the TRILEGAL-predicted color-magnitude density of foreground sources \textbf{towards each blank field}.  Note the difference in colorbar scale as compared to the left panel. \textbf{(c):} Total contamination by foreground Milky Way stars plus background galaxies from \textbf{panels (a) and (b)}
respectively, shown at a square root stretch to highlight density variations over the CMD.  The selection boxes identifying various subsamples used in our analyses are labeled.
\label{contamfig}}
\end{figure}

The resulting model for background galaxy contamination is shown in panel (b) of Fig.~\ref{contamfig}, where all recovered sources in the high latitude blank fields are overplotted and color coded by field as indicated in the upper right corner of the CMD.  The shading indicates the density of background galaxies, corrected for incompleteness using artificial star tests in each blank field, and we find a CMD distribution for background galaxy contaminants in good agreement with recent studies employing a similar approach \citep[e.g.][]{monachesi,m101}.

\begin{deluxetable}{lccclc}
\tablecaption{Blank ACS/WFC Fields Used to Assess Background Galaxy Contamination \label{blankfldtab}}
\tablehead{
\colhead{Field} & \colhead{Datasets} & \colhead{$t(F606W)$} & \colhead{$t(F814W)$} & \colhead{L} & \colhead{B} \\
\colhead{} & \colhead{} & \colhead{s} & \colhead{s} & \colhead{$^\circ$} & \colhead{$^\circ$} 
}
\startdata
CT344 & J8DD01010,J8DD01030 & 4080 & 4080 & 222.58 & -86.82 \\
ABELL-2744-HFFPAR & JC8N01010,JC8N01020 & 4846 & 4846 & 9.15  & -81.16  \\
MACS-J1149-HFFPAR & JCDUA2010,JCDUA2020 & 4981 & 4908 & 228.57  & 75.18  \\
ABELLS1063-HFFPAR & JCQT37010,GCQT45010 & 5046 & 5146 & 349.37 & -60.02 \\
GO-12608          & JBRB05010,JBRB06010 & 5250 & 5250 & 60.80 & 68.13 \\
GO-10874          & J9QT40030,J9QT40040 & 3619 & 3580 & 256.65 & 48.54 \\ 
\enddata
\end{deluxetable}

The total CMD density of foreground Galactic contaminants plus background galaxy contaminants is shown in the upper right panel of Fig.~\ref{contamfig} for the AMIGA4 field (although TRILEGAL-predicted foreground contamination differs by $<$2\% between the two AMIGA fields due to their similar Galactic latitudes).  Note that the color scales for Milky Way versus background galaxy contaminants differ by approximately an order of magnitude, so that the color scale for the sum of the two is shown at a square root stretch in panel (c) to display its full range of variation over the CMD.  In both the RC+RGBB region and the HB region, background galaxies are more important as contaminants than foreground Milky Way stars, although they remain fairly insignificant in all but the AMIGA6 field: The total (foreground plus background) contamination fraction in the RC+RGB CMD region is $<$1\% in the Brown\_stream field, 2$\pm$1\% in the AMIGA4 field, and 18$\pm$8\% in the AMIGA6 field, and contamination fractions in the HB region are similar to within 1\%.

As an additional check on our estimates of foreground and background contamination, which rely heavily on Galaxy models, we also take a more conservative, empirical approach.  We select among all of the AMIGA parallel imaging fields the one (AMIGA12) at the highest Galactic 
latitude and process it identically to AMIGA4 and AMIGA6.  If we assume \textit{all} sources in this field are contaminants, we obtain total contamination fractions in the RC+RGB CMD region of 2$\pm$1\% and 13$\pm$8\% for AMIGA4 and AMIGA6 respectively, in good agreement with our previous analysis given large Poissonian uncertainties stemming from the sparsity of the AMIGA12 field.  

To summarize the contribution of foreground and background contaminants, it is apparent from Fig.~\ref{contamfig} that the LF of foreground contaminants decreases towards fainter apparent magnitudes, while the LF of background galaxy contaminants increases towards fainter apparent magnitudes.  However, we argue that the systematic impact of these contaminants on our measured RC properties is negligible.  This is due not only to the small contamination fractions in the RC+RGB CMD region, but because we allow the LF slope $B$ to float as a free parameter, the contribution of contaminants effectively gets absorbed into this coefficient.  This is directly evidenced by decreasing values of the best-fit value of the LF slope $B$ moving radially outward from M31 where predicted contamination fractions increase.  

\subsubsection{M31 Halo \label{halocontamsect}}
 
The Brown\_stream field has a contamination fraction of $\sim$23\% due to the M31 spheroid, based on a comparison of radial velocity histograms cleaned
of Milky Way foreground stars using a carefully honed set of photometric and spectroscopic diagnostics \citep{jason_gss_spec}.  However, without such an ensemble of diagnostic information at hand for the AMIGA fields, we may still compare their stellar density with the stellar density predicted by an ensemble of M31 halo fields to obtain a crude estimate of M31 halo contamination in the AMIGA GSS fields.  Specifically, the minor axis fields analyzed by \citet{brown_minoraxis} are combined with two halo fields at $R_{\text{proj}}$=59 and 66 kpc with similar Galactic latitude as our AMIGA fields and archival ACS/WFC $F606W,F814W$ imaging (GO-11658, PI:Turnshek) processed with \texttt{Dolphot} identically as the AMIGA fields.  To estimate the contamination fraction, we simply count stars in the same (foreground extinction corrected) RC+RGB CMD region used to build LFs.  We additionally restrict the magnitude range
used for the star counts to be within 0.5 mag faintward of the RC in $F814W$, as 
background galaxy contamination increases 
moving faintward from the RGBB, apparent in Fig.~\ref{contamfig}.  This additional restriction results in $\leq$0.5 field stars per arcmin$^2$ (based on the TRILEGAL model) and a density of background galaxies statistically equivalent to zero.

\begin{figure}
\plotone{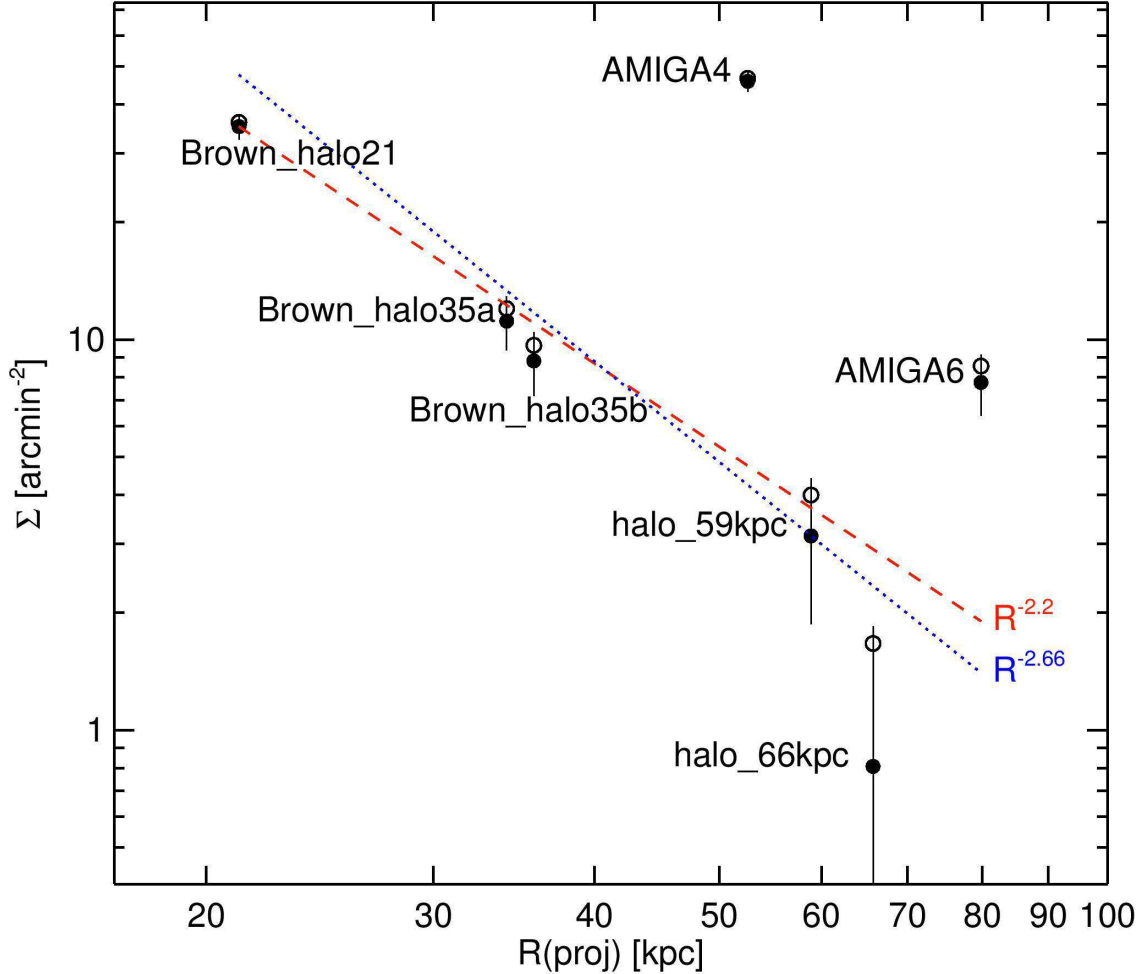}
\caption{Stellar density in several M31 halo fields measured from number counts in the RC region (see text for details).  Values before subtraction of Milky Way foreground stars and background galaxies are shown as open circles, and corrected values are shown as filled circles, where error bars represent the quadrature sum of (Poissonian) uncertainties in observed number counts, Milky Way foreground contribution, photometric errors, and background galaxy contribution.  For illustrative purposes, the power law projected density profiles with slopes of -2.2 from \citet{gilbert_sb_global} and -2.66 from \citet{ibataglobal} are shown as red dashed and blue dotted lines respectively, normalized to the fields which give the highest contamination fraction at the $R_{\text{proj}}$ of our AMIGA4 and AMIGA6 target fields. \label{halocontamfig}}
\end{figure}

In Fig.~\ref{halocontamfig}, we plot both the raw and the foreground- and background-decontaminated number counts in each field (where error bars include uncertainties on the number of observed stars as well as contaminants) versus projected distance from M31.  While it is clear qualitatively that the AMIGA4 and AMIGA6 fields are not dominated by the M31 halo, we quantify fractional contamination by the M31 halo and its uncertainty 
by assuming power law indices of -2.2$\pm$0.2 \citep{gilbert_sb_global} and -2.66$\pm$0.19 \citep{ibataglobal} for the projected M31 halo density profile.  For each power law index, the zeropoint is set by choosing the halo field yielding the highest (i.e.~most pessimistic) contamination fraction, and the corresponding slopes are illustrated in Fig.~\ref{halocontamfig}.  Accounting for uncertainties in the corrected star counts in each field as well as the quoted uncertainties on each power law slope, the more pessimistic of the two power laws gives  
an M31 halo contamination fraction of 10$\pm$2\% for the AMIGA4 field at 52 kpc, and 25$\pm$7\% for the AMIGA6 field at 80 kpc.  
This simplistic estimate neglects both the change in line-of-sight distance distribution of M31 halo stars with projected radius, as well as the $\sim$-0.01 dex per kpc metallicity gradient observed spectroscopically in the M31 halo (\citealt{gilbert_feh_global}, but see \citealt{brown_minoraxis}) as well as any age gradient which may be present.  However, it is also clear in Fig.~\ref{halocontamfig} that the observed (corrected) star counts are not entirely consistent with the power law slopes reported in the literature.  One simple explanation is that the halo fields we have employed actually have a non-zero contribution from additional M31 substructure. \citet{pandas_substructure} and \citet{ibataglobal} illustrate that most halo fields may coincide with substructure, and our chosen halo fields are no exception, e.g.~Stream D for the 35 kpc minor axis fields, and the SW cloud for the 59 and 66 kpc halo fields.  Meanwhile, a power law fit to the star count data in the halo fields shown in Fig.~\ref{halocontamfig} gives a slope of -2.8$\pm$1.0, consistent with the aforementioned values especially in the presence of some substructure.  In summary, the AMIGA6 field at 80 kpc has an M31 halo contamination \textit{fraction} consistent with the 21 kpc Brown\_stream field within the RC+RGB CMD region, although the observed metallicity gradient in the halo \citep{gilbert_feh_global} predicts that the M31 halo contaminants should be $\sim$0.6 dex more metal-poor towards the AMIGA6 field.  On the other hand, the AMIGA4 field at 52 kpc provides a relatively clean GSS sample, particularly in RC+RGB CMD region, where the total contamination fraction from the M31 halo, the Milky Way and background galaxies is $<$15\%.

\subsubsection{M31 Disk}

The locations of the AMIGA fields correspond to a radius of $>$200kpc from Andromeda in the
plane of the disk, 
so contamination from
Andromeda's disk is expected to be negligible ($<<$1\%).  Similarly, disk contamination
is expected to be $<$1\% in the halo fields along the minor axis of M31 \citep{brown_streamvsph}, and the
null detection of disk contaminants in the velocity distributions of \textit{all} fields
over the range of projected radii considered here by \citet{gilbert_feh_global} 
lends strong justification to our disregard of a potential M31 disk component 
contributing to the stellar populations in any of the fields considered.    

\section{Discussion \label{discusssect}}

\subsection{The Geometry of the GSS}

The observed mean magnitude and width of the RC may both be used together to constrain the geometry of the GSS.  In order to minimize systematic uncertainties, 
we define two
quantities in a strictly differential sense, and we use them to evaluate the difference between each of the AMIGA fields (at 52 and 80 kpc) versus the Brown\_stream field at 21 kpc. 

\subsubsection{Line of Sight Distance}
 
The first quantity $\Delta$$D = D(\rm{AMIGA})-D(\rm{Brown\_stream})$ represents 
the relative line-of-sight distance (in kpc) 
between each of the AMIGA fields and the Brown\_stream field.  In the left panel of Fig.~\ref{derivedparamsfig}, we show histograms of $\Delta$$D$ over all MCMC iterations, color coded by field.  Based on our mean RC magnitudes in $F814W_{0}$, the fields at 52 and 80 kpc both lie 
more than 50 kpc 
behind the 
Brown\_stream field, and in the case of the 52 kpc AMIGA4 field, our maximum likelihood fits constrain this line of sight distance to better than 
8\% relative precision.  However, in the sparse AMIGA6 field at 80 kpc, number statistics limit relative precision to $\sim$40\%, although they exclude a null distance gradient at 97.5\% confidence.

\subsubsection{Line of Sight Distance Scatter \label{rcmagdispersionsect}}

Second, the quantity $\Delta$$\sigma$$d_{los}$ denotes the excess line of sight distance scatter in the AMIGA fields compared to the Brown\_stream field.  This quantity is calculated by making the assumption \citep[e.g.][]{rc_geodist,nataf_los} that the observed $F814W$ width of the RC $\sigma$$I(RC)$ results from the (quadrature) sum of its intrinsic width $\sigma_{int}$, photometric errors $\sigma_{phot}$, and its geometrical line-of-sight broadening $\sigma_{geo}$:  

\begin{equation}
\sigma I(RC)^{2} = \sigma_{int,I}^{2} + \sigma_{phot,I}^{2} + \sigma_{geo}^{2}
\label{eqsigigeo}
\end{equation}

And correspondingly in $F606W$:

\begin{equation}
\sigma V(RC)^{2} = \sigma_{int,V}^{2} + \sigma_{phot,V}^{2} + \sigma_{geo}^{2}
\label{eqsigvgeo}
\end{equation}

However, we have the additional constraint that line of sight distance scatter will affect both filters equally, 
so that $\sigma_{geo}$ is equal in Eqs.~\ref{eqsigigeo} and \ref{eqsigvgeo}.  This allows fixing $\sigma_{int,V} / \sigma_{int,I}$ for any chosen value of $\sigma_{int,I}$, so with  
$\sigma_{phot,V}$ and $\sigma_{phot,I}$ at the RC magnitude in hand from the artificial star tests, 
we use the distance between fields $\Delta$$D$ (measured from $I(RC)$ above) to convert the difference in $\sigma_{geo}$ between fields $\sigma_{geo}(\rm{AMIGA})-\sigma_{geo}(\rm{Brown\_stream})$ in magnitudes to a line of sight distance difference in kpc, $\Delta$$\sigma$$d_{los}=\sigma$$d_{los}(\rm{AMIGA})-\sigma$$d_{los}(\rm{Brown\_stream})$.  The problem still remains that we must assume a value for $\sigma_{int,I}$, so we calculate $\Delta$$\sigma$$d_{los}$ under two extreme assumptions for $\sigma_{int,I}$ to evaluate their impact.  

\begin{figure}
\plotone{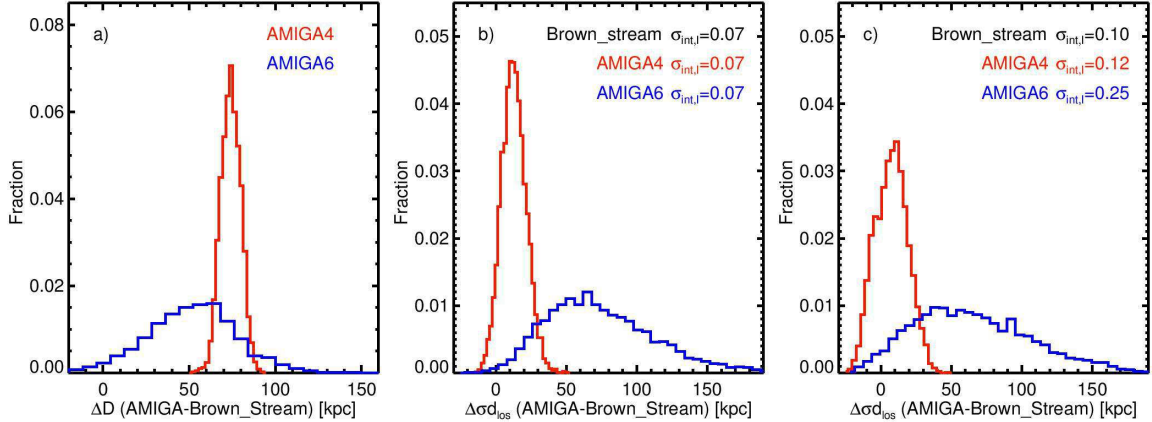}
\caption{Derived parameters from our unbinned maximum likelihood fits, color-coded by field.  \textbf{(a):} Difference in line-of-sight distance between each AMIGA field and the 21 kpc Brown\_stream field, calculated from the difference in $I(RC)$.  
\textbf{(b):} Difference in line-of-sight distance scatter between each AMIGA field and the 21 kpc field, calculated assuming an intrinsic RC dispersion of 0.07 mag.  \textbf{(c):} Same, but assuming the maximum intrinsic RC dispersion allowed in each field, given in the upper right corner \textbf{see text for details}. \label{derivedparamsfig}}
\end{figure}

\begin{enumerate}

\item First, we assume that the intrinsic width of the RC in each field is $\sigma_{int,I}$=0.07 mag, essentially the smallest allowed value once photometric errors are taken into account given the value of 0.065 ascertained from a single-age, single-metallicity population, the Galactic globular cluster 47 Tuc \citep{nataf_sigint}.  The resulting values of $\Delta$$\sigma$$d_{los}$, representing the increase in 1$\sigma$ line-of-sight distance scatter in each AMIGA field compared to the 21 kpc Brown\_stream field, are shown in the middle panel of Fig.~\ref{derivedparamsfig}.  In this case, where we have assumed that essentially all of the broadening of the observed $\sigma I(RC)$ is due to line of sight effects, both the AMIGA4 and AMIGA6 fields show an increased line of sight distance scatter at a moderate significance of 1.6 and 2.2$\sigma$ respectively.   

\item We make the opposite assumption, namely that the vast majority of the contribution to the observed RC width $I (RC)$ is intrinsic, and set $\sigma_{int,I}^{2}$=$\sigma I(RC)^{2} - \sigma_{phot,I}^{2}$ for each field.  The results, plotted in the right panel of Fig.~\ref{derivedparamsfig}, illustrate that the increase in $\sigma_{int}$ results in values of $\Delta$$\sigma$$d_{los}$ which are now compatible with zero for the AMIGA4 field.  Meanwhile, the AMIGA6 field is relatively unaffected, still showing an increased distance scatter at 1.7$\sigma$.    

\end{enumerate}

By requiring a consistent $\sigma_{geo}$ in both filters, we find values of 1.06$<$$\sigma_{int,V} / \sigma_{int,I} $$<$1.44 for both the Brown\_stream and AMIGA4 fields, in good agreement with evidence from both observations and theory \citep[e.g.][]{correnti_rc,girardirc} that the width of the RC in $V$ will generally be somewhat larger than in $I$ over a broad range of age-metallicity relations.  However, for the AMIGA6 field, strong evidence for relatively large values of $\sigma_{int}$ (the second of the two cases described above) is found in the unphysical values of $\sigma_{int,V} / \sigma_{int,I}$$<$0.5 required to obtain a constant $\sigma_{geo}$ unless the vast majority of the contribution to $\sigma I(RC)$ comes from $\sigma_{int,I}$ ($\sigma_{int,I}$$\gtrsim$0.2).  
One clear source of such a difference in $\sigma_{int,I}$ without invoking any difference in metallicity is an increase in age spread, and we now explore this issue in more detail.

\subsection{The Red Clump as a Standard Candle \label{standardcandlesect}}

We have thus far implicitly assumed that the RC is a standard candle, at least in a relative sense among our fields.  However, this ceases to be a valid assumption if any of our fields have substantially different star formation histories.
The similarity in both RC peak color (Sect.~\ref{rccolorsect}), HB morphology (Sect.~\ref{hbcolsect}) and photometric metallicity (Sect.~\ref{photmetsect}) across our target fields argue against bulk differences in metallicity distribution, but the RC luminosity even at fixed metallicity is not entirely insensitive to age.  Fortunately, this sensitivity remains at $\leq$0.03 mag per Gyr \citep[e.g.][]{gs01,girardirc} for ages above a few Gyr, which seems to be reasonably representative of the GSS population at least at $R_{\text{proj}}$$\lesssim$20 kpc \citep{brownsfh,bernard_sfh}.  This implies that, for example, a shift of $\sim$2 Gyr in the mean age of the metal rich population from 21 kpc to 52 kpc would affect the distance gradient across our fields at the level of $\lesssim$30\%.  Unfortunately such an age difference would require deeper imaging to detect given the photometric errors and galaxy contamination in the AMIGA fields.  However, the mean age of 8-9 Gyr found for the 21 kpc GSS population \citep{brownsfh,bernard_sfh} allows us to at least place limits on the line of sight distance gradient among our fields since a population:  At one extreme, a population with an age close to a Hubble time would have $I(RC)$ only $\sim$0.08 mag fainter at solar metallicity according to both BaSTI and PARSEC \citep{parsec} models.  At the opposite extreme, the properties of the RC change drastically for ages $\lesssim$1.8 Gyr \citep{girardi_spurious,girardirc} whereas no such change is observed in the RC morphology between the 21 kpc and 52 kpc field.  Therefore, even if the metal-rich GSS population in the 52 kpc field were to have a \textit{bulk} age shift down to 2 Gyr, $I(RC)$ would shift brightward by $\sim$0.13 mag at solar metallicity, implying a distance gradient $\sim$65\% steeper than we find by assuming similar age distributions.  However, because the Brown\_stream field already has a non-negligible fraction of young metal-rich stars \citep{brownsfh}, this would be an extreme case, and there are three qualitative arguments against such a large fraction of metal-rich stars with ages $\leq$2 Gyr in the AMIGA4 field:
   
\begin{enumerate}

\item{To check for any indications of a difference in age distribution between the Brown\_stream and AMIGA4 fields, the difference between the two fields (after foreground and background decontamination) is shown in Fig.~\ref{diffhess}.  
The differenced CMD was calculated 
by shifting the (halo decontaminated) Brown\_stream field faintward by 0.2 mag in $F814W$, scaling it to the same density as the AMIGA4 field, propagating in quadrature the uncertainties in the total contaminant density and Poisson noise on the observed number of stars, so that the differenced CMD could be plotted in terms of total uncertainty. 
For comparison, we overplot scaled solar Dartmouth 
isochrones at ages of 2, 6 and 10 Gyr, illustrating that there is no strong evidence for a \textit{relative} difference in the fraction of young metal-rich stars between the 21 and 52 kpc fields.  Reassuringly, the differences in HB morphology quantified in Sect.~\ref{hbcolsect} are indeed seen in the differenced CMD, at appropriately moderate statistical significance, implying that a bulk difference in age distribution, if present, should be detectable.

The comparison to the isochrones also illustrates the need for deeper imaging to make any quantitative statements about the age distribution in the AMIGA fields, since below the 50\% completeness limit (shown as a grey dotted line in Fig.~\ref{diffhess}), not only do completeness corrections themselves become more uncertain, but the implicit assumption that incompleteness of background galaxy contaminants scales identically as the incompleteness of actual stars may break down.}   

\begin{figure}
\gridline{\fig{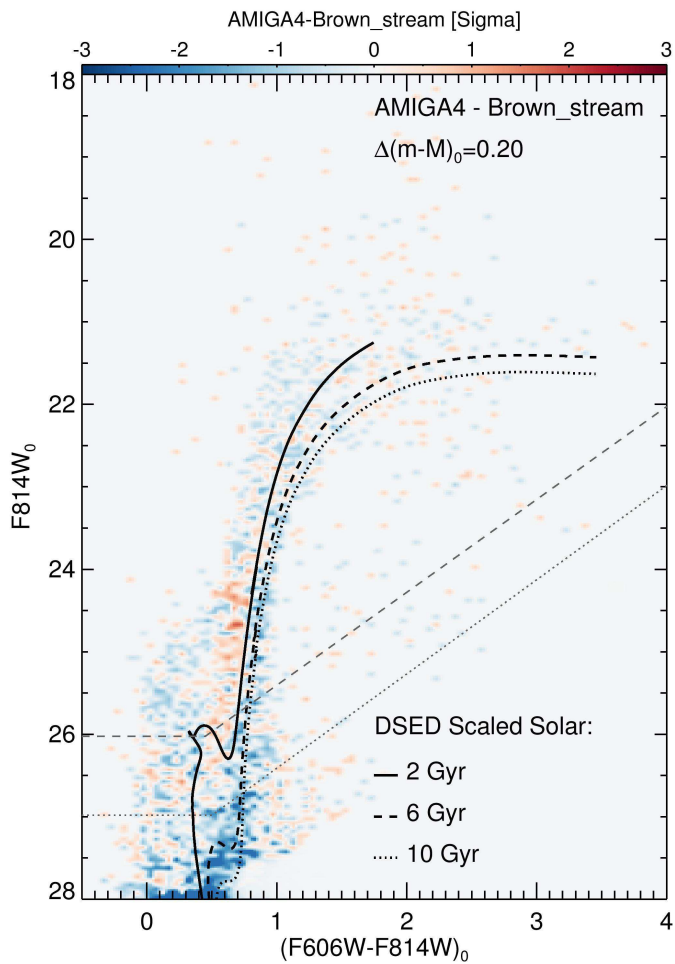}{0.50\textwidth}{}}
\caption{CMD illustrating the difference between the 21 kpc Brown\_stream field and the 52 kpc AMIGA4 field (see text for details), color-coded by statistical significance.  Scaled solar Dartmouth isochrones are overplotted at ages of 2, 6 and 10 Gyr, and the 90\% and 50\% completeness limits are shown as dashed and dotted lines respectively. \label{diffhess}}
\end{figure}

\item{Focusing on the upper RGB region used to derive photometric metallicities, the isochrones in Fig.~\ref{diffhess} illustrate that RGB color becomes increasingly sensitive to age at young ages.  In the present case, this well-known age-metallicity degeneracy operates 
such that if the metal-rich population of the AMIGA4 field had a much younger age than the metal-rich population of the 21 kpc Brown\_stream field, even higher actual metallicities would be required to yield the 10-Gyr-based photometric metallicities we calculated in Sect.~\ref{photmetsect}.   
Such a scenario would require that the AMIGA4 field has a \textit{mean} metallicity of at least solar, and while not implausible astrophysically, such high metallicities \textit{in the mean} are not seen among either the photometric metallicity maps of  \citet{ibataglobal}, the photometric metallicities along the GSS core from \citet{conn}, or the foreground-cleaned spectroscopic samples of \citet{gilbert_gss}.}

\item{Because the core helium burning evolution timescale of RC stars is a function of their mass, the RC itself can provide some clues on the mean age of the underlying metal-rich population via its strength relative to the RGB, $EW_{RC}$.  This is directly evidenced by a comparison between $EW_{RC}$ for the 11 kpc minor axis halo field and the 21 kpc stream field, which have similar mean $[\rm{Fe/H}]$ to within $\sim$0.1 dex but mean ages differing by $\sim$1 Gyr \citep{brownsfh,bernard_sfh}.  Accordingly, our maximum likelihood fits give a smaller $EW_{RC}$ for the older minor axis field at a statistically significant level, 1.036$^{+0.064}_{-0.059}$ versus 1.367$^{+0.098}_{-0.100}$ for the 21 kpc stream field. 

The similarity of $EW_{RC}$ (to well within their 1$\sigma$ uncertainties) across all of our target fields argues strongly against a bulk difference in age distribution, especially in the case of the 21 vs. 52 kpc fields where the uncertainties on both $EW_{RC}$ and metallicity are smaller.}

\end{enumerate}

In summary, the quality of our photometry does not allow quantitative comparisons between the age distributions of the AMIGA fields and the Brown\_stream field.  However, examining limiting cases at both age extremes demonstrates that the line of sight distance gradient between the 21 kpc and 52 kpc fields based on their $I(RC)$ values \textit{cannot be completely reconciled by an age difference}.

\subsection{Comparison to Recent Studies \label{litcompsect}}

\begin{figure}
\plotone{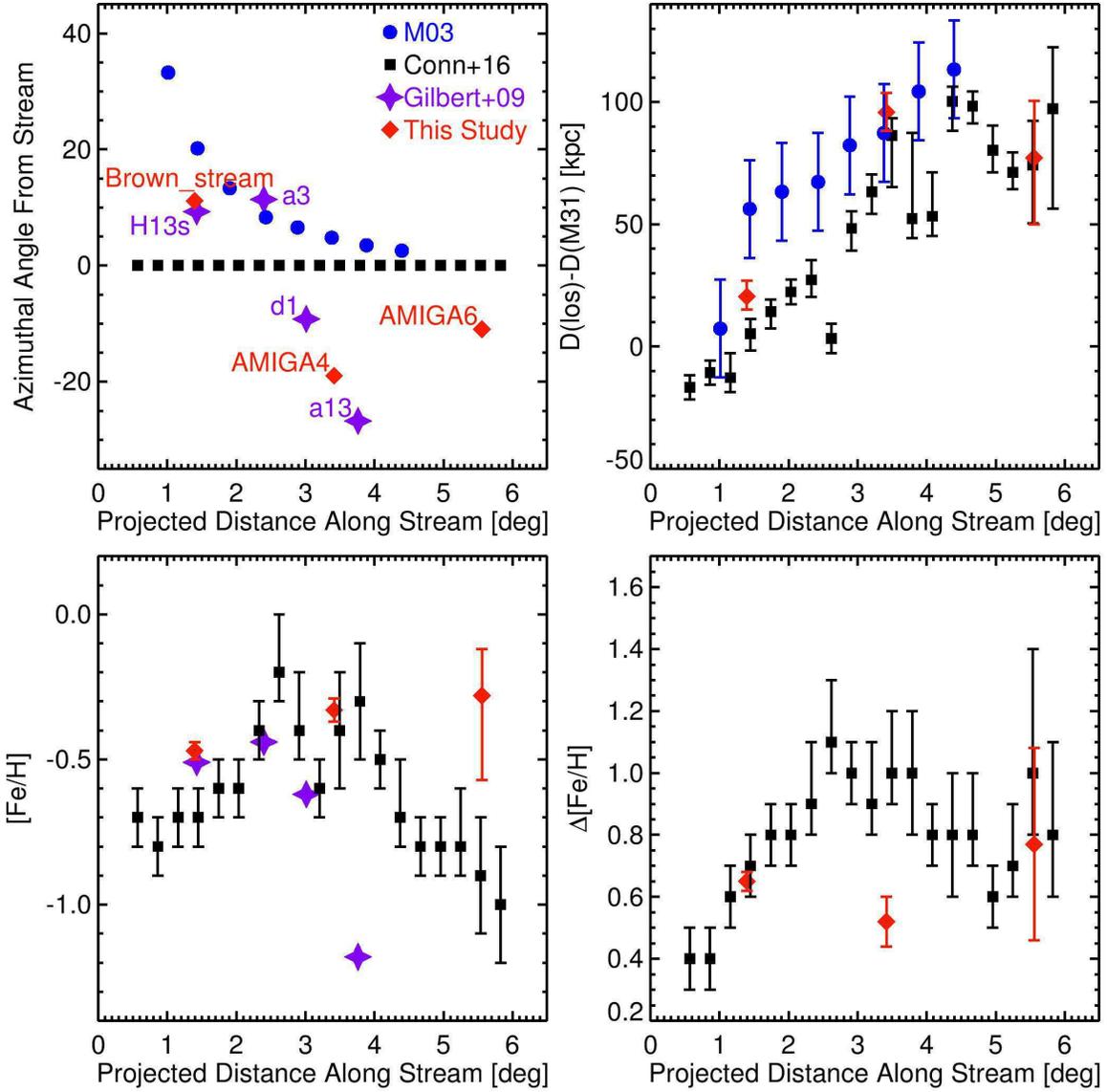}
\caption{Comparison between our results and other recent studies, in terms of projected distance on the sky along the GSS.  \textbf{Upper Left:} Map of our fields, showing projected distance along the GSS on the horizontal axis, where we have defined the GSS as a line passing through the \citet{conn} fields.  A projected distance of zero then corresponds to the intersection of this line with the major axis of M31.  On the vertical axis is the azimuthal angle from this line in degrees.  Our target fields are labeled and shown as red diamonds, the \citet{m03} fields M1-M8 are shown as blue circles, spectroscopic fields from \citet{gilbert_gss} are labeled and shown as magenta stars, and the \citet{conn} fields GSS1-GSS10 are shown as black squares.  \textbf{Upper Right:} Line of sight distance behind M31 as a function of projected distance along the GSS.  Distances for our target fields and their uncertainties are based on our $I(RC)$ measured from maximum likelihood fits in Sect.~\ref{rcmagsect}.  \textbf{Lower Left:} Median photometric metallicity from Sect.~\ref{photmetsect} compared to photometric metallicities from \citet{conn} and median spectroscopic metallicities from cleaned GSS samples presented in \citet{gilbert_gss}.  \textbf{Lower Right:} The width of the photometric metallicity distributions $\sigma$$[\rm{Fe/H}]$. \label{connfig}}
\end{figure}

In Fig.~\ref{connfig}, we plot several quantities for our target fields as a function of projected distance along the GSS in order to compare our results to other recent photometric and spectroscopic studies.  The projected distance along the GSS was computed by arbitrarily assuming that a line connecting the \citet{conn} fields demarcates the center of the GSS, and taking the intersection of this line with the major axis of M31 as the zeropoint of projected distance along the GSS.  In the upper left panel of Fig.~\ref{connfig}, we plot a map with the azimuthal angle of our target fields (red diamonds) from this line as a function of distance along the GSS, along with fields GSS1-10 from \citet{conn} (black squares) which have an azimuthal angle of zero by definition, as well as fields M1-M8 from \citet{m03} and several spectroscopic fields in the vicinity of the GSS from \citet{gilbert_gss} which are labeled and discussed below.  

In the upper right panel of \ref{connfig}, we plot the line-of-sight distance beyond M31 for our target fields, compared to TRGB measurements from \citet{m03} and \citet{conn}.  For consistency with \citet{conn}, these distances were calculated assuming $(m-M)_{0,M31}$=24.44, and in order to convert our \textit{relative} distance measurements from the RC mean magnitude to \textit{heliocentric} distances as they measured, we propagate in quadrature their uncertainties on their measurement of the distance to M31.  Lastly, our $\Delta$$D$ values were converted to total line-of-sight distances assuming that the 11 kpc minor axis field from \citet{brown_streamvsph,browncats} is at the distance of M31 (any changes in this assumption would only result in small zeropoint shifts in the plot).  
The distance gradient along the GSS implied by the difference in $I(RC)$ between the 21, 52 and 80 kpc fields is 
in excellent agreement with the results of \citet{conn} from TRGB measurements along the GSS core.  
Our target fields appear to be slightly more distant, 
although such a small difference ($\sim$5 kpc) could simply be due to differences in color-$T_{eff}$ transformations since their observations used different filters, and we note that we have used an identical isochrone set and assumed M31 distance to minimize systematics between our results and those of \citet{conn}.  

In the lower two panels of Fig.~\ref{connfig}, we plot the median $[\rm{Fe/H}]$ (lower left panel) and its 1$\sigma$ width $\sigma$$[\rm{Fe/H}]$ (lower right panel), where our photometric metallicities from Sect.~\ref{photmetsect} are compared with photometric metallicities from \citet{conn} and foreground-cleaned spectroscopic metallicities from \citet{gilbert_gss}.  
Again, the trend of $[\rm{Fe/H}]$ along the GSS in fig.~5 of \citet{conn} is reproduced well in a relative sense, since they also find an initial increase in $[\rm{Fe/H}]$ of $\sim$0.2-0.3 dex moving away from M31 along the GSS out to $R_{\text{proj}}$$\sim$45 kpc.  Our median $[\rm{Fe/H}]$ of -0.47$\pm$0.02 for the 21 kpc Brown\_stream field is slightly higher than the \citet{conn} value, and initially seems at odds with the mean $[\rm{Fe/H}]$=-0.7 measured by \citet{brownsfh} using the same imaging.  However, this difference can be largely attributed to their higher assumed foreground extinction of $E(B-V)$=0.08 from then-recent reddening maps (and a slightly shorter M31 distance), and if we instead use their values, we obtain a median $[\rm{Fe/H}]$=-0.67$\pm$0.04, in much better agreement with their value from maximum likelihood fits to isochrones.  Our value is also in good agreement with the median $[\rm{Fe/H}]$=-0.51 measured spectroscopically from a cleaned sample in the co-located H13s field \citep{gilbert_gss} and a median $[\rm{Fe/H}]$=-0.37 from a comparison to BaSTI models \citep{bernard_sfh}.  Also, while our photometric metallicitity for the 80 kpc AMIGA6 field appears somewhat ($<$2$\sigma$) higher than the \citet{conn} value, recall that our upper-RGB-based values for this field were extremely sensitive to the assumed foreground contamination model (see Sect.~\ref{photmetsect}).  Furthermore, the difference in RC peak color from Sect.~\ref{rccolorsect} implies a best fit value of -0.64 dex and a 1$\sigma$ lower limit of $\sim$-1 dex for this field, in good agreement with \citet{conn}.  Similarly, despite large uncertainties, the difference in $I^{RC}_{RGBB}$ between the 52 and 80 kpc fields argues for a decrease of 0.2$\pm$0.3 dex in metallicity according to the relation of \citet[][see their sect.~6]{natafrgbb}.       
In addition, the median $[\rm{Fe/H}]$ we measure for all three fields are in good agreement with the photometric metallicity maps of \citet{ibataglobal}, especially since a larger line-of-sight distance for our fields means their $[\rm{Fe/H}]$ estimates may run low by up to $\sim$0.2 dex.

There are two primary differences between the trends observed for our target fields and those in the recent literature.  
The first difference between the AMIGA fields westward of the GSS core and what is found by \citet{conn} along the GSS core is in the \textit{width} of the metallicity distribution.  The lower left panel of Fig.~\ref{connfig} illustrates that while our measured MDF widths in the 21 and 80 kpc fields are in good agreement with the \citet{conn} results, we find a significantly smaller width for the AMIGA4 field at 52 kpc.  This is not due to their use of the slightly higher \citet{sfd} foreground reddening values, since this only affected our MDF widths at $\sim$0.02 dex.      
Since the \citet{conn} fields lie on the GSS core, broadening of the metallicity distribution by M31 halo contaminants should play a larger role for our AMIGA fields, potentially worsening the discrepancy.  However, 
the decreasing $[\rm{Fe/H}]$ width with increasing $R_{\text{proj}}$ we observe from the 21 kpc to 52 kpc fields is in excellent agreement with spectroscopic values from carefully cleaned GSS core samples out to $\sim$30 kpc analyzed by \citet[][see their fig.~15]{gilbert_gss}.  Furthermore, \citet{conn} performed their fitting assuming that the MDFs for the GSS component in each of their fields was Gaussian, while Fig.~\ref{photmetcmdfig} illustrates that this is likely not the case, and it is unclear what the impact of such an assumption would be.

The second difference between our results and those in the literature can be used to improve constraints on \textit{azimuthal} metallicity gradients perpendicular to the GSS.    Leveraging our values together with the ensemble of GSS fields studied by \citet{gilbert_gss}, they measure values of $[\rm{Fe/H}]$ which are significantly ($\gtrsim$0.6 dex) more metal poor than either their GSS core fields or our AMIGA fields for field a13, which lies in the GSS envelope westward of the AMIGA4 field.  Meanwhile, higher metallicities consistent with values found in the GSS core are reported for the d1 field which lies eastward of the GSS envelope.  The high $[\rm{Fe/H}]$ we measure for the AMIGA4 field (more metal-rich than the \citealt{gilbert_gss} value but in excellent agreement with the \citet{conn} value) then restricts the location of a dropoff in metallicity to be westward of the AMIGA4 field but eastward of field a13, and indeed hints of such a dropoff are seen in fig.~8 of \citet{ibataglobal} running north-south at $\xi$$\sim$0.

We also briefly mention model-to-model differences in the context of our results shown in Fig.~\ref{connfig}.  We have used the same isochrones and assumed M31 distance as \citet{conn} in an attempt to minimize such systematics, although an earlier version of this draft employed Victoria-Regina isochrones and a Galactic [$\alpha$/Fe]-$[\rm{Fe/H}]$ law.  We found that this choice had a negligible impact on our measured RC parameters, since the location of the RC+RGB CMD region was only slightly affected.  This is because its vertical (magnitude) boundaries were chosen empirically from our rough-guess RC apparent magnitude, and its boundaries in color move only slightly since model-to-model differences in predicted $(F606W-F814W)_{0}$ color are small at the RC luminosity compared to the gradient of color with metallicity.  However, this is not necessarily the case on the upper RGB, where differences in stellar atmosphere models, synthetic spectra, and color-$T_{eff}$ transformations are more apparent with increasing metallicity and decreasing surface gravity, and model-to-model differences primarily manifested as substantial ($\sim$0.4 dex) zeropoint shifts in $\Delta$$[\rm{Fe/H}]$ in the lower left panel of Fig.~\ref{connfig}, warning that while the \textit{relative} gradients we measure along the GSS seem to be insensitive to the choice of evolutionary models, the \textit{absolute} width of the MDFs we measure are indeed dependent on the choice of model and [$\alpha$/Fe].    

This study essentially serves as a demonstration case for using the RC to place constraints on distance, age and metallicity, and suffers from two primary limitations.  The first, number statistics (in total and relative to stellar and galaxy contaminants), are a limiting factor due to the size of the ACS/WFC and WFC3/UVIS field of view.  This problem is alleviated via wide-field ground-based imaging and/or spectroscopic campaigns, which thus far have lacked the depth to probe the RC/HB and therefore rely on the TRGB and upper giant branch \citep{gilbert_feh_global,ibataglobal,conn}.  Meanwhile, it has been demonstrated that the RC of M31 and the GSS is well within the reach of current ground-based imagers on large telescopes, allowing constraints from the TRGB and RC to be leveraged together \citep[e.g.][]{tanaka}, ideally using additional filters to remove foreground and/or background contaminants \citep{m31hypersuprimecam}.  
The second limitation inherent to our observations is their photometric depth, preventing us from using the main sequence turnoff and subgiant branch to disentangle age and distance effects.  This problem has been overcome by using deeper space-based imaging to probe small fields down to their main sequences, allowing the use of models to construct star formation histories \citep{brown_minoraxis,browncats,bernard_sfh}.  The SFH can be recovered (albeit with decreased precision) using observations such as ours which reach the RC but not the MSTO \citep[e.g.][]{dolphin_sfh}, but these approaches are generally applied to stellar populations such as dwarf galaxies which are known or assumed to have fractionally small line-of-sight distance distributions.  While this may be a reasonable assumption for the aforementioned studies targeting fields within a few tens of kpc in $R_{\text{proj}}$, our results as well as previous observations \citep{m03,conn} and models \citep{kirihara,m31_recentsims} imply that it is likely not the case over the full extent of the GSS.

Our success in the use of the RC as a standard candle implies that it can be a valuable discriminant between major and minor merger scenarios.  For example, \citet{m31_recentsims} make specific, testable predictions for the line-of-sight distributions of GSS debris, and our data indicate that in the presence of adequate number statistics, such distributions can be quantified and compared to model predictions in detail.  In the well-populated 21 kpc field, the formal uncertainty on $I(RC)$ of $\pm$0.005 mag translates to a relative distance uncertainty of $\sim$2 kpc, sufficient to disentangle a multi-modal line-of-sight distance distribution of metal-rich debris.

\section{Conclusions}

We have used an unbinned maximum likelihood technique to characterize the red clump towards three fields along the GSS at projected distances of 21, 52 and 80 kpc.  By comparing the fields observationally in a strictly differential sense, we avoid assumptions specific to any set of stellar models or the true underlying star formation history of any of our fields.  By measuring the red clump magnitude and its dispersion in our three target fields, we find:

\enumerate

\item The distance gradient across fields azimuthally offset from the GSS core is in excellent agreement with the distance gradient measured along the GSS core by \citet{conn}, arguing against any dramatic change in line of sight distance moving azimuthally away from the GSS.  
The exact steepness of this distance gradient could be significantly affected by presently undetectable differences in the age distribution of the metal rich GSS stars in our target fields, but it cannot be negated.

\item The width of the line-of-sight distance distribution likely increases with projected distance from M31.  In the 52 kpc AMIGA4 field, this result could be plausibly nullified by an increased age spread for the metal-rich GSS population as compared to the 21 kpc field.  However, in the 80 kpc AMIGA6 field, the increase in line of sight distance distribution is significant at 1.7$\sigma$ even in the presence of a larger age and/or metallicity spread causing an intrinsic RC width of up to 0.25 mag in $F814W$.

\item Based on the both the color of the RC peak and the metallicity distribution measured photometrically, the metallicity in the 52 kpc AMIGA4 field is at least as high as the GSS core at 21 kpc.  Together with a notably more metal-poor metallicity distribution found westward of the AMIGA4 field at similar $R_{\text{proj}}$, this implies a dropoff in median metallicity moving westward located at $\xi$$\sim$0 (the Right Ascension of M31), hinted at in existing photometric metallicity maps based on the upper RGB.

\acknowledgements

Support for Program GO-14268 (Project AMIGA) was provided by NASA through a grant from the Space Telescope Science Institute, which is operated by the Association of Universities for Research in Astronomy, Inc., under NASA contract NAS 5-26555. 

\vspace{5mm}
\facilities{HST (ACS/WFC,WFC3/UVIS)}


\appendix

\begin{figure}
\gridline{\fig{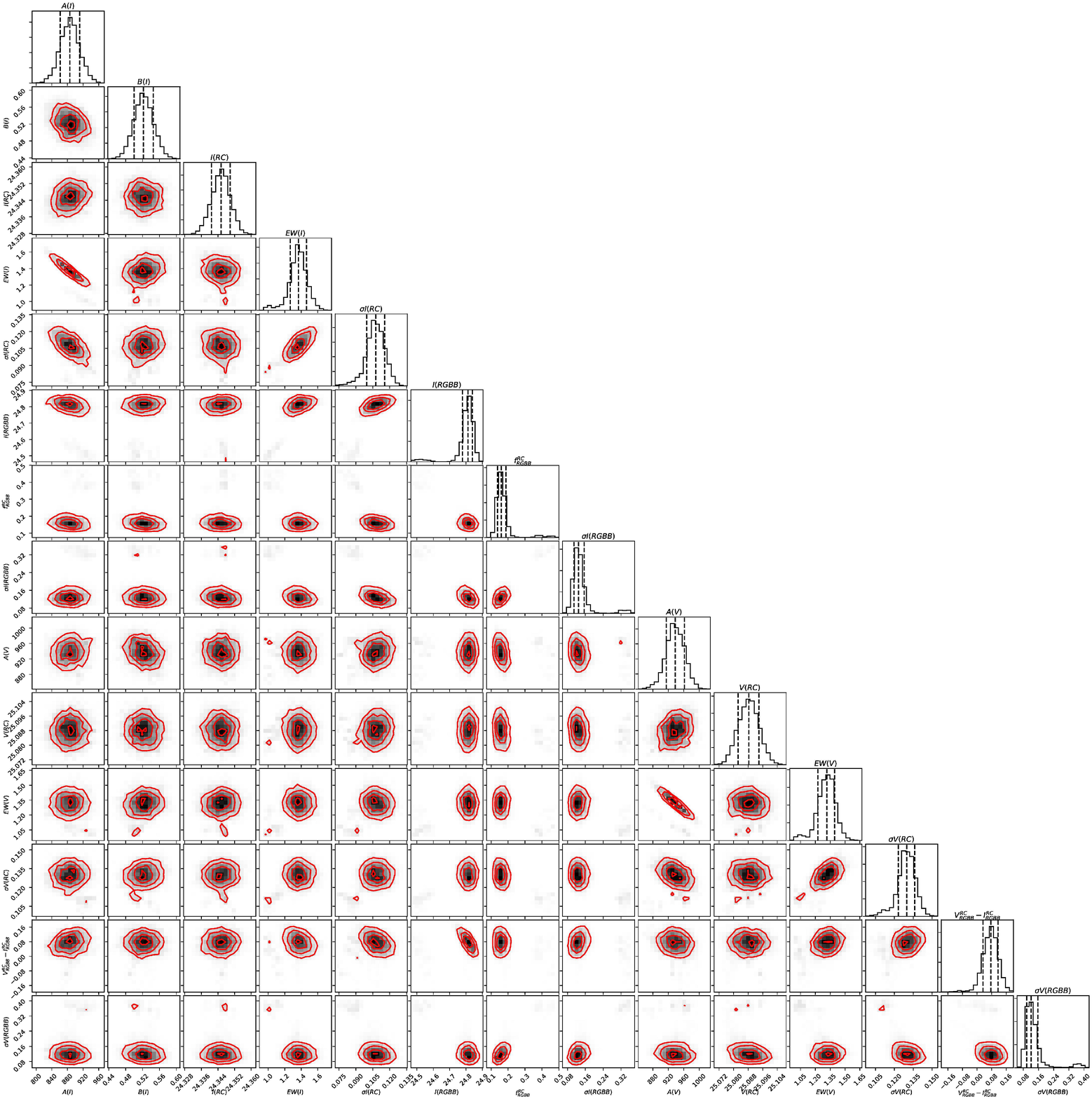}{0.99\textwidth}{}
	 } 
\caption{Corner plot illustrating the posterior distribution functions for the LF parameters in the Brown\_stream field and their interdependence, taken directly from the MCMC runs described in Sect.~\ref{rcmagsect}.  For each parameter, the histogram at the top of the corresponding column shows the 68\% confidence interval enclosed within vertical dotted lines, and the two-dimensional plots illustrating correlations between parameters include [0.5,1,1.5,2]$\sigma$ contours shown using red lines. \label{cornerbrownstreamfig}}
\end{figure}

\begin{figure}
\gridline{\fig{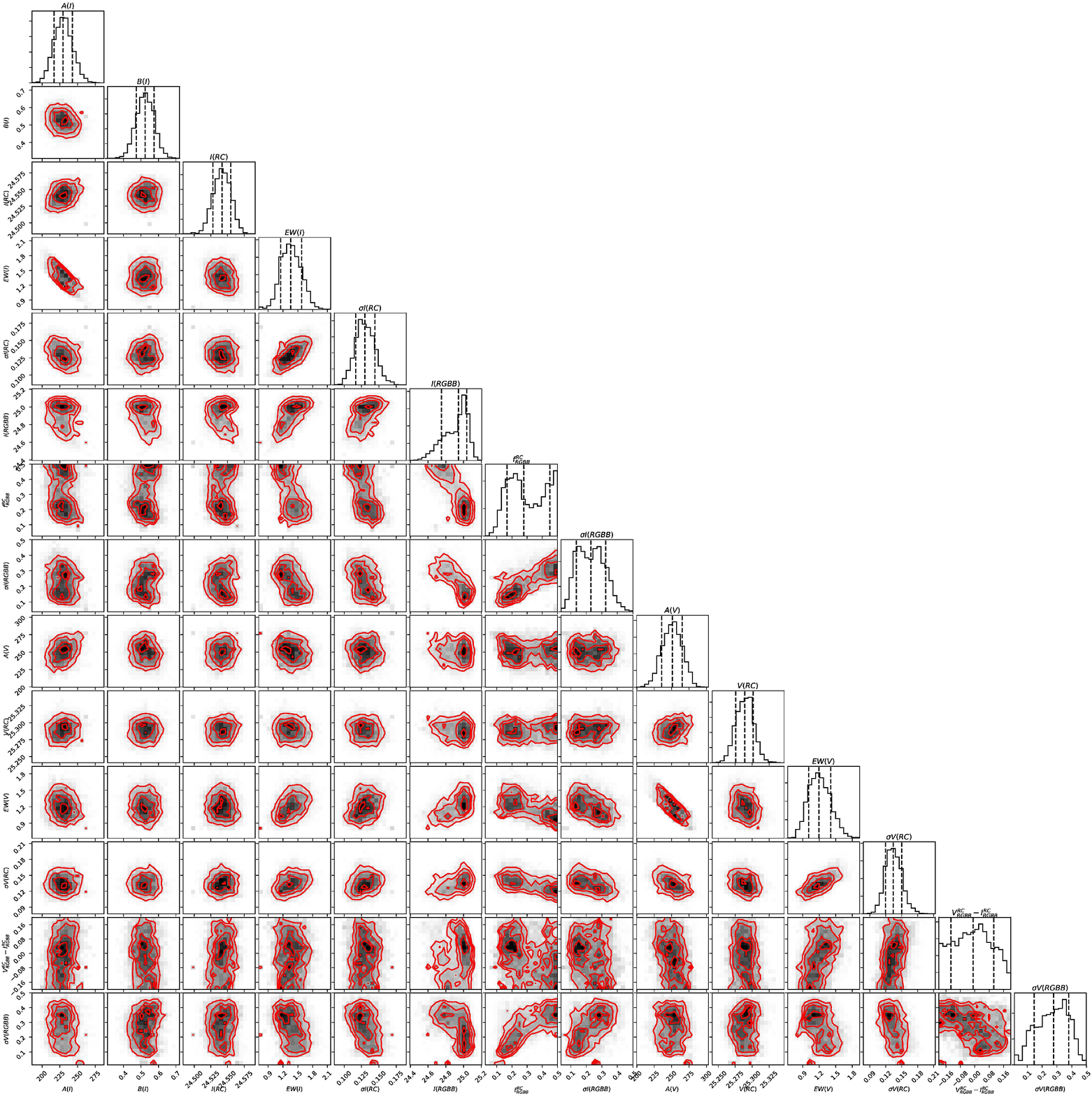}{0.99\textwidth}{}
	 }
\caption{As in Fig.~\ref{cornerbrownstreamfig} but for the AMIGA4 field. \label{corneramiga4fig}}
\end{figure}

\begin{figure}
\gridline{\fig{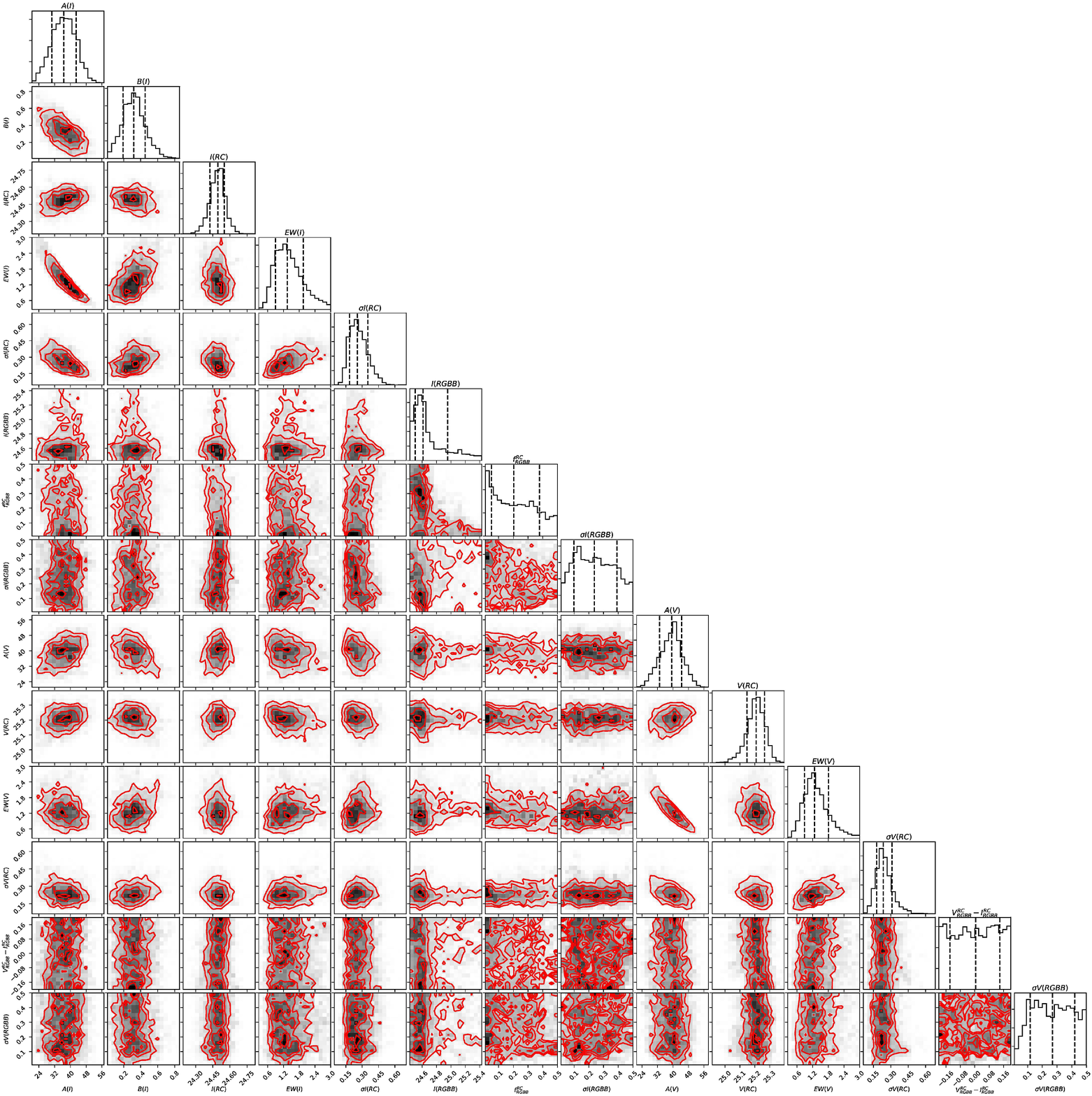}{0.99\textwidth}{}
	 }
\caption{As in Fig.~\ref{cornerbrownstreamfig} but for the AMIGA6 field. \label{corneramiga6fig}}
\end{figure}

\end{document}